\documentclass[apjl]{emulateapj}

\usepackage{amsmath}
\usepackage{amssymb}
\usepackage{amsfonts}
\usepackage{amstext}
\usepackage{graphicx}
\usepackage{fancybox}
\usepackage[usenames,dvipsnames]{color}
\usepackage{pstricks}
\usepackage{bm}
\usepackage{soul, color}
\usepackage{latexsym}
\usepackage{pifont}
\usepackage{shorttoc}
\usepackage{subfigure}
\usepackage{textcomp} 
\usepackage{float}
\usepackage{enumitem}
\usepackage{longtable}

\usepackage{bm}
\usepackage{lipsum}
\usepackage{threeparttablex}
\usepackage{hyperref}
\hypersetup{urlcolor=cyan}

\def\aap{A\&A}

\def\apjs{ApJS}
\def\apjl{ApJL}

\newcommand{\angs}{\, {\rm \AA}}

\newcommand{\no}[1]{}%per comentar espais grans
\renewcommand{\exp}[1]{\mathrm{exp}\left(#1\right)}

\newcommand{\myemail}{lluis.mas-ribas@jpl.nasa.gov}
\def\lsim{~\rlap{$<$}{\lower 1.0ex\hbox{$\sim$}}}

\def\gsim{~\rlap{$>$}{\lower 1.0ex\hbox{$\sim$}}}

\shorttitle{Radiation-Pressure Waves and Multiphase Quasar Outflows}
\shortauthors{Ll. Mas-Ribas}

\begin{document}

\title{Radiation-Pressure waves and  multiphase Quasar Outflows}

\author{Llu\'is Mas-Ribas\altaffilmark{1,2,3}} 
\altaffiltext{1}{Jet Propulsion Laboratory, California Institute of Technology, 4800 Oak Grove Drive, 
Pasadena, CA 91109, U.S.A.\\
\url{\myemail}}
\altaffiltext{2}{California Institute of Technology, 1200 E. California Blvd, Pasadena, CA 91125, U.S.A.}
\altaffiltext{3}{Institute of Theoretical Astrophysics, University of Oslo,
Postboks 1029, 0315 Oslo, Norway.
\\
\\
\\
\copyright 2018. All rights reserved.}

%-------------------------------ABSTRACT---------------------------
\begin{abstract}  

       We report on quasar outflow properties revealed by analyzing  more than 60 
composite outflow spectra built from $\sim 60\,000$ C{\sc iv} absorption troughs in the 
SDSS-III/BOSS DR12QBAL catalog. We assess the dependences of the equivalent widths of 
many outflow metal absorption features on outflow velocity, trough width and position, and quasar 
magnitude and redshift. The evolution of the equivalent widths of the O{\sc vi} and N{\sc v} lines  
with outflow velocity correlates with that of the mean absorption-line width, the outflow electron density, 
and the strength of lines arising from collisionally-excited meta-stable states. None of these correlations  
is found for the other high- or low-ionization species, and different behaviors  with trough width 
are also suggested.  We find no dependence on quasar magnitude or redshift in any case.  All the 
observed trends can be reconciled by considering a multiphase stratified outflow structure, where 
inner regions are colder, denser and host lower-ionization species. Given the prevalence of 
radiative acceleration in quasar outflows found by \cite{Masribas2019}, we suggest that radiation pressure 
sweeps up and compresses the outflowing gas outwards, creating waves or filaments where the 
multiphase stratified structure could take form. This scenario is supported by  
the suggested correlation between electron density and outflow velocity, and the similar behavior 
observed 
for the line and line-locking components of the absorption features. We show that this outflow 
structure is also consistent with other X-ray, radiative transfer, and polarization results, 
and discuss the implications of our findings for future observational and numerical quasar outflow 
studies.

\end{abstract}

%-------------------------------INTRODUCTION----------------------------------------------------------------
\section{Introduction}

      A fraction of about $20-30\,\%$ of quasars  show in their spectra broad 
absorption lines (BALs) of high-ionization species on the blue side of the corresponding quasar 
emission lines  \citep{Hewett2003,Reichard2003,Ganguly2008,Knigge2008,Allen2011}.  
Because these broad absorption troughs can partially mask the quasar emission lines, 
they are assumed to trace outflows outside the quasar broad line region (BLR), where the emission 
lines are formed. Specifically, these outflows are thought to consist in the material that was orbiting 
the central supermassive black hole and that has been accelerated outwards 
\citep[e.g.,][]{Murray1995,Elvis2000,Proga2000,Proga2004}. 

    BALs appear blueshifted from the emission 
sources by velocities of up to $10-20\,\%$ of the speed of light 
\citep{Weymann1981,Trump2006,Gibson2009,Rodriguez2011}, and about $20\,\%$ of them are 
detached from the corresponding emission lines \citep{Korista1993}. In most cases, only broad features 
of high-ionization species are observed (HiBALs; \citealt{Weymann1991,Filizak2013}), 
but a fraction of about $15\,\%$ of outflows \citep{Sprayberry1992} also show broad absorption lines 
of low ions \citep[LoBALs;][]{Voit1993,Gibson2009}.   

     Studies of BALs indicate that quasar outflows generally consist of large column densities of 
highly-ionized material, with $\log (N_{\rm H}/{\rm cm^{-2}})\gtrsim 22-23$ \citep[e.g.,][]{Moravec2017}, 
and super-solar metallicities \citep{Hamann1998}. Since the absorption lines are often saturated but not 
completely black, it is believed that the outflows cover the emission sources partially, the exact value 
depending on the ionic species and line of sight \citep[][see also \citealt{Proga2012} for such a behavior 
in disk winds]{Hamann1993,Arav1999}. However, the bottom 
of the absorption troughs could also be (additionally) filled with radiation from other directions that has 
been scattered by the outflow into the line of sight \citep{Lee1997,Baek2007}. Although it is clear 
that these outflows likely reside outside the BLR, their extent 
is still uncertain by more than two orders of magnitude, from $\lesssim1$ pc to hundreds of pc 
\citep{Arav2018,Hamann2018}. 

    Quasar outflows have typically been detected and studied through the analysis of BALs at  
ultraviolet (UV) frequencies in the quasar rest frame \citep{Lynds1967}. Nowadays, however, the 
outflow absorption signatures show a wide range of complex profile shapes and widths, 
and they are detected in both UV and X-ray. Specifically, BALs refer to absorption troughs 
blueshifted by more than $3\,000\,{\rm km\,s^{-1}}$ from the respective quasar emission line, and 
with a full width at half {\it minimum} larger than $\sim 2\,000\,{\rm km\,s^{-1}}$ \citep{Weymann1991}. 
When the features are narrower, down to a few hundreds of ${\rm km\,s^{-1}}$, but one can still 
`decide' that they correspond to quasar outflows, these adopt the name of mini-BALs, and below 
this poorly-defined threshold the troughs are named narrow absorption lines 
\citep[NALs; e.g.,][]{Hamann2004}. This classification arises from differences in the properties 
measured for these absorbers but it is possible that they all trace the same type of outflows and 
the differences are driven by orientation, time of observation, or spatial location within the outflow 
\citep[see][and references therein]{Hamann2018}. We will refer to all the absorbers in this work as 
troughs, and will analyze them independently of the aforementioned nomenclature.  We will show 
that the distinct type of absorbers do indeed trace different media within the outflow, and that they 
arise naturally when considering the evolution of the equivalent widths of hydrogen and metal lines 
with outflow parameters.

    In this paper, we aim to gain insight into the physical properties of quasar outflows represented 
by absorption troughs via the analysis of composite outflow spectra. By doing so, we will 
neither reveal properties of the individual objects, which are known to vary broadly, nor their time 
dependence, which can be of the order of from days to years or more 
\citep[e.g.,][]{Proga2012,He2017,Misawa2018,Rogerson2018}. Instead, we will obtain the 
time-averaged mean properties of the outflow population and 
will be able to analyze the dependences of these properties on several parameters in a general 
fashion. Furthermore, the stacking of many spectra will result in 
the reduction of the average spectral 
noise and will enable us to detect faint absorption features associated to the outflows. This  
is especially important in the spectral region of the Lyman-alpha forest, where the high density of 
intervening absorption lines overlaps with the outflow features making the detection of the latter 
difficult even in high-resolution spectra. In particular, current analyses that make use of the 
Lyman-alpha forest often have to discard the whole spectrum of quasars containing BALs ($\sim 
10\%$ of the spectra), as the outflow contamination in this region is unknown \citep[e.g., 
the cross-correlation studies by][or future work by \citealt{Desi2016}]{Busca2013,Fontribera2013,
Bautista2017,Perezrafols2018b,Perezrafols2018}. In these cases, our composite spectra can be  
used as absorption templates to mask the regions affected by the outflow,  
thus enabling the use of the uncontaminated spectral regions. 

    A number of works have analyzed composite outflow spectra 
\citep[e.g.,][]{Baskin2013,Baskin2015,Hamann2018} but they have mostly focused on 
correlations between the outflows and the emission sources. 
We present here a complementary approach; our analysis focuses on correlations 
and dependences between physical parameters describing the absorption troughs, such as their 
position in velocity space, which we assume to trace the velocity of the outflow, their width, and 
the degree of detachment between troughs and sources, together with the dependences on  
quasar redshift and brightness. 

   In this work we build up on our previous paper, \cite{Masribas2019}, where 
we constructed a number of the stacks that we will analyze here, and where we studied the 
observational signature of radiative acceleration known as line locking \citep{Milne1926,Scargle1973,Arav1994}. 
In that work we found that the mean absorption lines  in the 
outflow composite spectra present an absorption feature on their blue side, which we identified 
as the C{\sc iv} line-locking component. Line locking was visible in all our stacked spectra, 
suggesting that radiation pressure is a prevalent mechanism for the acceleration of outflows. 
However, we found no evidence for the presence of line locking from other species, such as 
O{\sc vi}, N{\sc v} or Si{\sc iv}. We will consider the two components of each 
absorption transition (line and line locking) separately in our current analysis, and will investigate 
whether the properties of the outflows reveal the cause of our findings in the previous paper. 

      In \S~\ref{sec:spectra} below, we present the composite spectra used for our analysis, and  
detail the methods for their computation and analysis in \S~\ref{sec:methods}. 
The overall composite outflow spectra are presented in \S~\ref{sec:dissection}, and we analyze the 
dependences of the equivalent widths on quasar and outflow parameters in \S~\ref{sec:dependencies}. 
We propose a general structure for the outflow in order to explain our measurements in 
\S~\ref{sec:bump}, and discuss our findings and future work in \S~\ref{sec:discussion}, before 
concluding in \S~\ref{sec:conclusions}.

We assume a flat $\Lambda$CDM cosmology with the parameter values from \cite{Planck2015}.

\begin{table}\center    % - - - - -- - - - - - SUBSAMPLES - -  --  - --  - - - - - -
	\begin{center}
	\caption{Outflow samples}	\label{ta:subsample}
	\begin{threeparttable}
		\begin{tabular}{lrcr} 
		\hline 
		 Selection criteria		     &Mean		&~ $n\,{\rm (\AA)}$\,\tnote{($1$)} $\;$   &No. troughs	 \\ 
         all 									&			&$0.437\pm0.029$	&$59\,872$ \\ \hline
         
		\multicolumn{4}{c}{Outflow velocity (${\rm km\,s^{-1}}$)}	     \\ 
        $v  < 200$   	             &$139$	     &$0.335\pm0.064$	 &$185$      \\
        $v  < 350$                   &$264$	     &$0.331\pm0.045$	 &$3\,167$     \\
        $v  < 650$   	             &$315$	     &$0.254\pm0.040$	 &$4\,355$     \\
        $350 \leq v  < 650$          &$452$	     &$0.278\pm0.038$	 &$1\,188$     \\
        $650 \leq v  < 1\,500$         &$1\,122$   &$0.291\pm0.037$	 &$2\,278$     \\
        $1\,500 \leq v  < 3\,000$        &$2\,233$   &$0.387\pm0.081$	 &$6\,706$     \\
        $3\,000 \leq v  < 5\,000$        &$3\,980$   &$0.444\pm0.095$	 &$7\,727$     \\
        $5\,000 \leq v  < 8\,000$        &$6\,426$   &$0.462\pm0.031$	 &$10\,183$  \\
        $8\,000 \leq v  < 13\,000$     &$10\,330$  &$0.439\pm0.050$	 &$12\,488$  \\
        $13\,000 \leq v  < 17\,500$  &$15\,208$	 &$0.354\pm0.078$	 &$9\,075$     \\
        $v > 17\,500$   			 &$20\,867$	 &$0.302\pm0.138$	 &$10\,737$  \\ \hline
        
		\multicolumn{4}{c}{Trough width (${\rm km\,s^{-1}}$)}			   \\ 
        $v  < 560$   				&$508$	   &$0.227\pm0.045$	 &$11\,250$    \\
        $560 \leq v  < 708$  		&$629$	   &$0.264\pm0.094$	 &$8\,420$       \\
        $708 \leq v  < 1\,260$   	&$951$	   &$0.305\pm0.149$	 &$13\,488$    \\
        $1\,260 \leq v  < 2\,240$   &$1\,704$  &$0.381\pm0.057$	 &$10\,266$    \\
        $v  \geq 2\,240$   		&$5\,179$  &$0.560\pm0.056$	 &$20\,125$    \\ \hline

		\multicolumn{4}{c}{Trough min. velocity (${\rm km\,s^{-1}}$)}		\\ 
        $v  < 300$   			&$128$		&$0.423\pm0.048$	&$2\,574$   \\
        $300 \leq v  < 2\,300$  	&$1\,236$		&$0.431\pm0.075$	&$10\,713$  \\
        $2\,300 \leq v  < 5\,300$   &$3\,694$		&$0.381\pm0.132$	&$10\,661$  \\
        $5\,300 \leq v < 9\,000$   	&$7\,113$		&$0.352\pm0.032$	&$10\,356$  \\
        $9\,000 \leq v < 13\,500$   &$11\,183$	&$0.400\pm0.082$	&$10\,570$  \\
        $13\,500 \leq v < 18\,500$  &$15\,953$	&$0.409\pm0.054$	&$10\,059$  \\
        $v  \geq 18\,500$   		&$21\,186$	&$0.463\pm0.089$	&$8\,630$   \\ \hline

		\multicolumn{4}{c}{Quasar redshift}		\\ 
        $z  < 1.95$   			&$1.77$	&$0.449\pm0.093$	&$10\,580$             \\
        $1.95 \leq z  < 2.20$  	&$2.09$	&$0.455\pm0.047$	&$10\,584$             \\
        $2.20 \leq z  < 2.37$   	&$2.28$	&$0.455\pm0.042$	&$10\,812$             \\
        $2.37 \leq z < 2.60$   	&$2.48$	&$0.414\pm0.066$	&$10\,591$             \\
        $2.60 \leq z < 3.00$   	&$2.79$	&$0.452\pm0.009$	&$10\,880$             \\
        $z  \geq 3.00$   		&$3.38$	&$0.461\pm0.159$	&$10\,098$             \\ \hline

		\multicolumn{4}{c}{Quasar magnitude (mag)}		\\ 
        $-25.0 \leq M_i  < -22.0$  	&$-24.6$	&$0.522\pm0.191$	&$10\,972$    \\
        $-25.5 \leq M_i  < -25.0$  	&$-25.3$	&$0.482\pm0.100$	&$10\,912$    \\
        $-26.0 \leq M_i  < -25.5$   &$-25.8$	&$0.428\pm0.061$	&$12\,932$    \\
        $-26.5 \leq M_i < -26.0$   	&$-26.2$	&$0.415\pm0.040$	&$12\,020$    \\
        $-27.0 \leq M_i < -26.5$   	&$-26.7$	&$0.387\pm0.037$	&$8\,949$     \\
        $-30.0  \leq M_i < -27.0$   &$-27.5$	&$0.383\pm0.074$	&$7\,749$     \\ \hline
		
		\end{tabular}	
		\begin{tablenotes}
			\item[($1$)] {This parameter denotes the values of the intercept in Eq.~\ref{eq:ap} 
			and is an indicator of the width of the absorption lines in the composite spectra. }
		\end{tablenotes}
	\end{threeparttable}
	\end{center}
\end{table}

%-------------------      DATA     -----------------
\section{Quasar Outflow Data}\label{sec:spectra}

     We use the composite outflow absorption spectra of \cite{Masribas2019}, built from $\approx 60\,000$ 
broad ($>450\, {\rm km\,s^{-1}}$) C{\sc iv} absorption troughs identified in the quasar spectra of the 
twelfth data release of the SDSS-III/BOSS \citep{Eisenstein2011,Dawson2013} quasar catalog 
\citep[DR12Q;][]{Paris2016}.  We refer to the catalog of C{\sc iv} troughs as broad absorption line 
(BAL) catalog for consistency with the nomenclature in \cite{Paris2016}, DR12QBAL, although not all 
the troughs are BALs  strictly considering the definitions by, e.g., \cite{Weymann1991,Hall2002,
Trump2006}.  The DR12QBAL catalog quotes the parameters of the absorption troughs when these 
have widths of $>450\, {\rm km\,s^{-1}}$ with normalized quasar flux (i.e., the transmission defined by 
\citealt{Paris2012}) below  $0.9$ \citep{Paris2016}.

     In detail, we analyze the 36 outflow spectra described in Table \ref{ta:subsample}, which  
is the same as Table 2 in \cite{Masribas2019} and that we replicate  
here for clarity. These spectra are created from subsets of the total BAL catalog 
regarding outflow velocities (derived from the distance between the position of the minimum 
flux within each C{\sc iv} absorption trough and the corresponding C{\sc iv} emission line), 
velocity width of the troughs, degree of detachment between the 
troughs and the quasars (characterized by the minimum velocity of the troughs), and visual  
redshift and absolute $i$-band magnitude of the quasars.  All the parameter values are taken 
from the DR12QBAL catalog for the case of the troughs, and from the DR12Q catalog for the case of 
the quasar redshift and absolute magnitude, where further descriptions on their measurements are 
detailed. We initially consider all the  troughs in the DR12QBAL catalog for completeness, and do not 
apply any cut regarding the methods for which they were measured. Since our analysis concerns the 
comparison between ionic species and/or subsamples, the method for measuring parameters 
should not have a significant impact in our 
results and conclusions. The number of subsamples for each parameter is generally set to be the 
minimum in order to contain a large number of troughs (yielding high signal-to-noise 
composites), while enabling us to trace with precision the evolution of the line equivalent widths 
addressed in later sections. For instance, we create several subsamples with small numbers of 
objects at outflow velocities below $\approx 1\,000\, {\rm km\,s^{-1}}$ because, as we will discuss 
later, these velocities potentially trace halo gas and not outflows, and the measured equivalent widths 
for the line and line-locking components show different behaviors in this range. Other groups 
using our spectra can find helpful subsamples that differentiate between these velocity ranges (e.g., 
separating NALs and mini-BALs). Also, when possible, we apply the cuts in the data aiming to have 
similar numbers of objects in each bin to aid the comparisons between subsamples.
For the case of trough width, we perform a cut at the peak of the width distribution  (${\approx 
700\,}{\rm km\,s^{-1}}$ in Figure 2 of \citealt{Masribas2019}) given its sharpness, which results in 
one of the width bins with a smaller number of troughs compared to the others, but without impact 
on the results. Due to the shape of the same distribution, the sample with the largest 
width range  ($v  \geq 2\,240\,{\rm km\,s^{-1}}$) covers a broad range of values and has a 
large number of troughs, 
but additional cuts do not yield extra information.  The samples of quasar redshift and magnitude 
are built considering similar number of objects per bin while evenly covering the parameter space. 
The {\it first column} in Table \ref{ta:subsample} denotes the cuts applied to create the 
corresponding subsamples, and the mean values of the respective parameters are quoted in the 
{\it second column}. The {\it third column} denotes the values of the intercept, which is 
related to the width of the absorption lines via the modeling of the absorption features 
described by Eq.~\ref{eq:fit}, and the number of spectra contributing to each subsample 
is presented in the {\it fourth column}. 
These spectra are publicly available for future outflow studies at 
\url{https://github.com/lluism/BALs}.

We also build  
and use 30 additional outflow composites, each arising from the combination of one range of trough 
width and one range of outflow velocity  from the following respective sets:  we consider 
five ranges of trough width, $(0,560)$, $[560,1\,000)$, $[1\,000,2\,500)$, $[2\,500,5\,000)$ and 
$[5\,000,25\,000]$, and six ranges of outflow velocity, $(0,800)$, $[800,2\,000)$, $[2\,000,5\,000)$, 
$[5\,000,10\,000)$, $[10\,000,14\,000)$ and $[14\,000,25\,000]$, where all the values have units of 
${\rm km\,s^{-1}}$. 

        The final set of outflow composites for our analysis thus consists of one spectrum built from  
all the  troughs and 65 composites from the aforementioned subsamples. We summarize 
below the steps for creating these composite spectra.

%-------------------      Methods     -----------------
\section{Methods}\label{sec:methods}

       We detail the methods that we use for  computing the composite outflow spectra 
in \S~\ref{sec:composite}, and for modeling the absorption-line profiles in \S~\ref{sec:profile}. 
We briefly describe these calculations here but refer the interested reader to \cite{Masribas2019} 
and \cite{Masribas2016c} for further details. In \S~\ref{sec:eqw} 
we present the calculations of the equivalent widths, and discuss measurements of the 
column densities in  \S~\ref{sec:N}. 

%-------------------      Composite     -----------------
\subsection{Composite Spectra Computation}\label{sec:composite}

        The composite spectra are computed as a weighted mean, where the weights are derived from the 
signal-to-noise ratio (S/N) of the individual spectra, and where we use a mean quasar spectrum  
to remove the intrinsic quasar profile and obtain absorption-only outflow spectra. 

    We start by describing the computation of the mean quasar spectrum. For this calculation we consider  
all the quasar spectra in the DR12Q quasar catalog, instead of only those containing broad absorption 
troughs to avoid residual absorption features induced by the overlapping absorbers 
\citep[see Figure 3 in][]{Masribas2019}.  We include here the quasars with outflow absorption troughs 
because they are on average more luminous than those without outflow signatures (e.g., median 
$i$-band absolute magnitude $M_i \sim -26.3$ mag for BAL quasars against $M_i \sim -25.6$ mag 
for the non-outflow ones; \citealt{Hamann2018}), and they will thus contribute in general to the mean 
spectrum with weaker emission lines due to the Baldwin effect \citep{Baldwin1977}. These quasars 
correspond to $\approx 9\%$ of the total number in the DR12Q catalog, and their average outflow 
absorption features imply a variation of $<3\%$ in the mean quasar flux in the worst cases (i.e., next to 
the O{\sc vi} and N{\sc v} emission lines in Figure 3 of Mas-Ribas \& Mauland). The equivalent widths 
will be impacted by a similar fraction, which is much smaller than the uncertainties that we will 
compute. Overall, the inclusion or not of BAL quasars in this computation has no significant impact on  
our conclusions.  

      Each quasar spectrum is shifted to the quasar rest frame, rebinned into a  
regular array of pixel width of  $1\, {\rm \AA}$, and normalized by dividing it by 
the mean flux within two wavelength windows, $1\,300<\lambda_{\rm r}/{\rm \AA}<1\,383$ and 
$1\,408<\lambda_{\rm r}/{\rm \AA}<1\,500$, where $\lambda_{\rm r}$ denotes the quasar rest-frame 
wavelength. The selection of these windows enables a high S/N for the absorption lines in the 
Lyman-alpha forest given their proximity to that region. They cover a total 
wavelength range large enough to obtain a reliable estimate of the mean flux, despite the potential 
contribution of the faint emission lines of O{\sc i}$\,\lambda1305$ and C{\sc ii}$\,\lambda1335$. 
Using these windows for the composite calculation could bias the equivalent width results as the 
spectra may contain the Si{\sc iv} and C{\sc iv} absorption troughs in them.  We 
have compared the eleven composite spectra for outflow velocity bins in Table 
\ref{ta:subsample} that arise from using these windows and using the window  $2\,600<\lambda_{\rm r}/{\rm \AA}<2\,750$ for both the mean quasar spectrum and the composite calculation. We observe 
relative flux variations of $20-30\%$ between the two calculations at the peak of the strongest absorption 
lines, i.e., Ly$\alpha$, N{\sc v}, C{\sc iv}, and of a few percent in the unabsorbed regions and weaker 
absorption lines. The default window calculation results in larger fluxes. More important, these 
differences do not depend on outflow velocity except for the subsample with outflow velocities 
$1\,500 \leq v  < 3\,000$, where the default window shows lower flux values by a similar amount. 
Overall, these differences do not impact the trends that we will observe below so we use the default 
windows throughout. We also compute the mean S/N in these windows, which we use to assign to 
each spectrum $j$ a weight as 
\begin{equation}
w_j = {{1}\over{{{\rm S/N}_j}^{-2} + \sigma^2}} ~, 
\end{equation}
where $\sigma=0.1$ is a limitation to the contribution from very high S/N spectra (see 
\citealt{Masribas2016c} for discussions on other values).  We then compute the 
weighted mean of these normalized spectra to obtain the final composite quasar spectrum. 

   The outflow composites are calculated in a similar way. Here, however, we only use the DR12QBAL  
spectra, and shift all them to the position of the respective absorbers, for which we use the position 
of the minimum flux within the absorption troughs as discussed in sections 3.2 and 6.1 in 
\cite{Masribas2019}. We rebin the spectra into a new regular 
array, this time of pixel width of $0.3\, {\rm \AA}$ to obtain the highest spectral resolution enabled 
by the BOSS spectra ($\approx 69\,{\rm km\,s^{-1}}$), and normalize each of them with the 
corresponding mean flux in the same two windows used for the mean quasar spectrum. Each normalized 
spectrum at the rest-frame position of the outflow (i.e., absorption trough) is then divided by 
the previously computed mean quasar spectrum, shifted also to the position of the absorber. This 
step eliminates the quasar-emission profile and results in an almost flat spectrum with the 
outflow absorption features. The final composites are computed as the weighted mean of 
the absorption spectra of interest for each subsample, using the weights that were assigned to each 
spectrum in the mean quasar calculation. 

%-------------------      line profile modeling      -----------------
\subsection{Line Profile Modeling}\label{sec:profile}

   The absorption line profiles are modeled considering the line and line-locking components of 
each of the transitions reported in Table 1 of \cite{Masribas2019}, and publicly available 
at \url{https://github.com/lluism/BALs}. 

    We first renormalize each composite spectrum since residual contamination still 
exists after the composite computation. We do so by dividing the spectrum by a pseudo-continuum,  
obtained by smoothing the absorption-free regions of the spectrum itself with a Gaussian kernel 
with $\sigma_{\rm G}=10$ pixels. Despite this renormalization, the unabsorbed parts of 
the outflow spectra can be not completely flat in some cases, so we model the profiles  
accounting for a local continuum computed over the individual absorption features. This local 
continuum is obtained as a linear fit to a number of pixels in both sides of each absorption feature 
(see section 3.3.1 in \citealt{Masribas2019} for further details on these 
procedures).

     Next, a least-squares fit is performed to the line and line-locking components of each transition 
within a wavelength window containing the absorption feature, and considering the expression 
\begin{align}\label{eq:fit}
F_\lambda = C_\lambda\, {\rm exp} & \left[  -b\, \exp{\frac{-(\lambda-\lambda_{\rm c})^2}{2a^2}} \right. 
\\ \nonumber
& ~ \left.  -d\, \exp{\frac{-(\lambda-\lambda_{\rm ll})^2}{2a^2}} \right] ~ .
\end{align}
Here, $C_\lambda$ denotes the local continuum, and $\lambda_{\rm c}$ and $\lambda_{\rm ll}$ are 
the line and line-locking positions, respectively. As we argued in \cite{Masribas2019}, we set  
a separation of $497\,{\rm km\,s^{-1}}$ between the line and line-locking components, equivalent to 
the distance between the two lines of the C{\sc iv} doublet. Given the resolution of the BOSS spectra, 
however, variations of $|\delta_{\rm v}|\lesssim 30\,{\rm km\,s^{-1}}$ in the relative position of the line 
locking do not change the fits significantly, but we adopt the C{\sc iv} value because it is physically 
motivated. We fit the absorption features considering a half-window size of 4 $\angs$ around  
the center of the absorption lines, and adding an extension of $500\,{\rm km\,s^{-1}}$ on the blue 
side to account for the line-locking component. In case of overlapping between multiple windows 
we consider a single window from the lowest to the highest limit of the intervening windows, and fit 
all the profiles simultaneously. In general, this yields to the fitting of up to five absorption lines, and 
up to five for the range covering the O{\sc vi} doublet. Two ranges at around $1125$ and $1195$ \angs, 
covering ten weak features each, are split in four ranges of five lines each. We have explicitly checked 
that these divisions do not impact the measurements. 
The parameters $b$ and $d$ in Eq.~\ref{eq:fit} represent the depth of the absorption features and 
are set free, while 
the parameter $a$ is related to the width of the lines and is fixed considering the spectral resolution 
of the BOSS spectra and its wavelength dependence. Specifically, we compute the mean values  
of the $a$ parameter obtained when this is set free in the various subsamples for each transition, and 
use these mean values to model the slope of the evolution of $a$ with wavelength (Figure 4 in 
\citealt{Masribas2019}). This calculation results in the linear fit 
\begin{equation}\label{eq:ap}
a  = (3.4\pm0.6)\times 10^{-4}\, \lambda   + n ~,
\end{equation}
where $n$ is the intercept obtained for each subsample by fitting the above equation to the free $a$ 
values of the strongest transitions in each case. The values and uncertainties of $n$ for each 
subsample are reported in the {\it third column} of Table \ref{ta:subsample}, and all parameters in the 
above equation have units of \angs. A few uncertainties for the $n$ parameter appear 
much larger than the others. These are produced by some $a$ parameters used for the estimation 
of $n$ that depart strongly from the average values. This likely reflects uncertainties in the fitting, 
as observed for the C{\sc iv}$\,\lambda1548$ line in some cases, where this line is overestimated 
and the line-locking C{\sc iv}$\,\lambda1550$ component underestimated. Details on the 
aforementioned calculations and parameters are further discussed in sections 3.3.2 and 3.3.3 
of \cite{Masribas2019}.

%-------------------        EQW  	calculation    -----------------
\subsection{Equivalent Width}\label{sec:eqw}

  The equivalent width of each line and each line-locking feature in all the composites is measured by 
integrating the modeled absorption profile under the continuum within the corresponding absorption 
window. In detail, we integrate $1-F_\lambda/C_\lambda$. 
  
%-------------------        EQW  	error    -----------------
\subsubsection{Equivalent Width Uncertainty}\label{sec:eqwerr}

        Although the bootstrap method is suitable for calculating the equivalent width uncertainties 
in stacking studies \citep{Masribas2016c}, the large number of spectra and samples 
addressed in the current work would require performing an unfeasible number of spectral stacks 
(of the order of $10^5$). Instead, we derive the uncertainty of the equivalent widths 
from the individual normalized outflow absorption spectra, following the method by 
\citealt{Arinyo2018} (their section 3.4). We emphasize that this method, as well as the bootstrap, 
takes into account the noise from the pixels but it does not include the systematic uncertainty from the 
modeling of the continua or the line profiles.

       The equivalent width uncertainty for an absorption feature in one single spectrum is computed as 
\citep[Eq. 2 in][]{Arinyo2018} 
\begin{equation}
\sigma_{\rm W} = \overline{\sigma}_{\rm t}\, N^{1/2}_{\rm p}\, {\rm d}\lambda ~,
\end{equation}
where ${\rm d}\lambda=0.3$ ${\rm \AA}$ is the wavelength step in the normalized outflow spectra, 
and $\overline{\sigma}^2_{\rm t}$ is the mean of the squared transmission error 
${\sigma}^2_{{\rm t}i} = \sigma^2_i/C^2_i$ of the $N_{\rm p}$  pixels  in the two ranges used for the 
calculation of the continuum. The calculation of the uncertainty is performed in these ranges 
instead of in the line windows to avoid possible effects due to correlations between neighboring pixels 
in the BOSS spectra \citep{Arinyo2018}. The terms $\sigma^2_i$ and $C^2_i$ in the previous expression 
denote the squared values of the uncertainty and continuum, respectively, for a pixel $i$.

    For each feature in a composite spectrum we then compute the mean of the squared 3-sigma 
clipped uncertainties of the spectra that contribute to that composite. The final equivalent width 
uncertainties are simply the square root of these means.

%-------------------      COLUMN DENSITIES     -----------------
\subsection{Column Density}\label{sec:N}

      In \cite{Masribas2016c} we developed a theoretical model for estimating the column 
densities from the equivalent width of the absorption features. This model is intended for the analysis 
of composite spectra and it takes into account the saturation of the absorption lines. These 
characteristics make it useful for our current work as absorption lines that are in reality 
saturated can appear weak in our composite spectra due to the BOSS resolution, 
and thus the equivalent width is the only quantity that can be directly measured from the stacks 
\citep{York2000}. 

     However, the outflow material often does not cover the totality of the radiation source 
(partial covering) and thus 
the absorption lines, although saturated, do show a residual flux at the bottom of the troughs 
\citep[e.g.,][]{Hamann1993,Hamann1998,Arav1999,Borguet2012}. When 
this happens the equivalent widths can lead to incorrect (underestimated) column density measurements   
and, consequently, unrealistic outflow parameters \citep[][and references therein]{Arav2013,Chamberlain2015}.
Furthermore, the outflow composites contain a small number of strong transitions of each species, and 
the uncertainties in their equivalent widths are typically large. These two factors prevent us from 
accurately calibrating the model, as this 
requires the largest possible number of transitions for (at least) one species \citep[see section 6 
in][]{Masribas2016c}. In turn, the inferred outflow column densities are 
highly sensitive to the choice of the range covered by the parameter space and the values 
assigned to the priors in our Monte Carlo computation. We therefore decide to not attempt the 
measurement of the column densities in this paper and defer it to future work where this can 
be addressed via numerical modeling.

%-------------------------------------- DISSECTION ---------------------------------------
\section{The Mean Outflow Absorption Spectra}\label{sec:dissection}

 \begin{figure*}\center %------------  BAL STACK  
\includegraphics[width=1\textwidth]{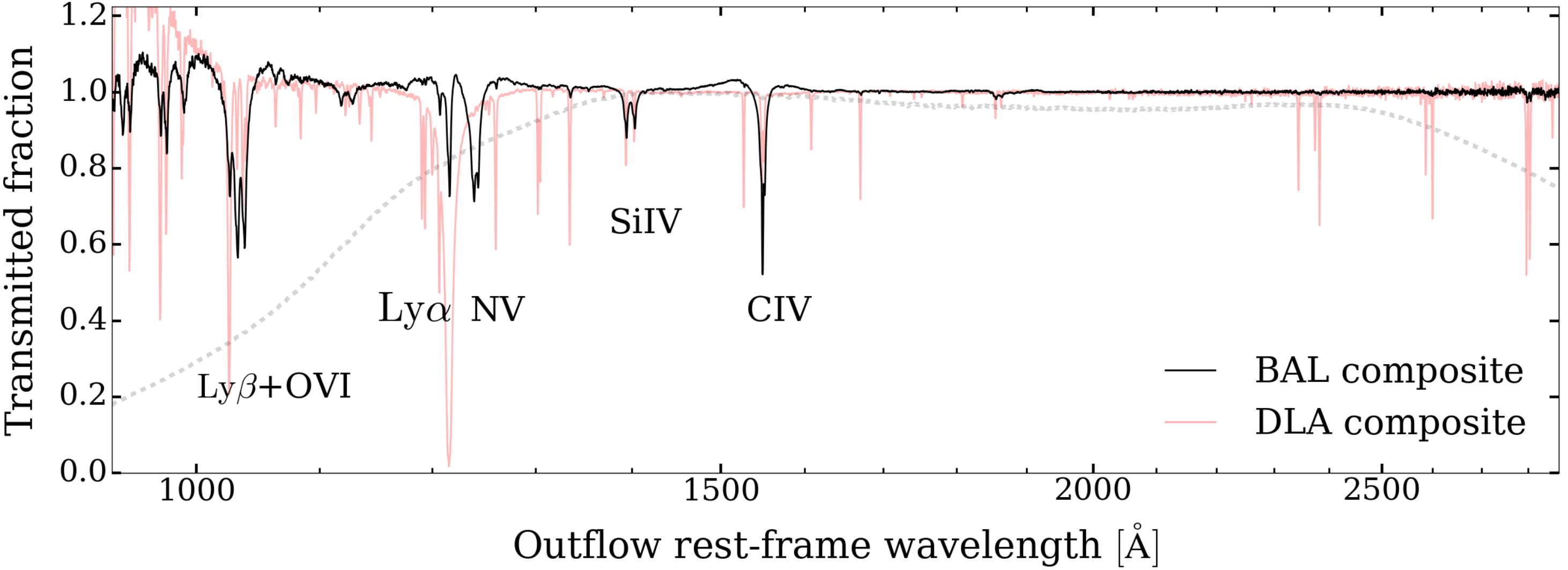}
\caption{Outflow absorption composite spectrum considering all the troughs in the 
BAL catalog ({\it black line}). The strong absorption lines of highly-ionized species 
such as C{\sc iv}, Si{\sc iv}, N{\sc v} and O{\sc vi} indicate a general 
high-ionization state for the outflowing medium, although low-ionization species  
are also visible, suggesting a multiphase structure. For comparison, the {\it pale red line} 
shows the composite spectrum of the `total sample'  of DLAs in \cite{Masribas2016c}. 
This spectrum describes systems with large hydrogen column densities associated to 
galaxies and thus reveals much weaker high-ionization absorption lines 
and stronger low-ionization features compared to that of quasar outflows. The outflow spectrum is 
normalized for visualization using a pseudo-continuum computed by smoothing the spectra 
with a Gaussian kernel of $\sigma_{\rm G}=80$ pixels that results in artificial flux 
transmission above the unity around some of the strongest lines. This  spectrum is  
therefore only adequate for visualization purposes. The faint {\it gray dashed line} denotes 
the fraction of spectra in the BAL catalog that are used for the outflow composite at 
every pixel, decreasing toward low and high wavelength values due to the redshift of 
the quasars and the spectral coverage of the BOSS spectrograph. The horizontal axis 
is logarithmically spaced.}
\label{fig:stack}
\end{figure*}

   Figure \ref{fig:stack} shows the outflow composite spectrum of all the C{\sc iv} 
troughs in the BAL catalog (`all' sample in Table \ref{ta:subsample}). 

  The {\it black line} in Figure \ref{fig:stack} displays the central wavelength region 
of the outflow spectrum, and the {\it pale red line} shows the final composite spectrum 
for the `total sample' of Damped Lyman-alpha (DLA) systems in \cite{Masribas2016c}   
detailed below. Both spectra are normalized for comparison, using a pseudo-continuum 
computed by smoothing the spectra with a Gaussian kernel of $\sigma_{\rm G}=80$ 
pixels. This simple smoothing introduces spurious flux features that rise above the 
unity next to some of the strongest absorption lines, and so these spectra are only 
valid for visualization purposes. The faint {\it gray dashed line} denotes the fraction of spectra in the 
BAL catalog that contributes to this composite at every pixel, decreasing toward low and high 
wavelengths due to the redshift of the quasars and the spectral coverage of the 
BOSS spectrograph. 

      Contrary to the case of quasar outflows that represent highly-ionized material on average, 
the low-ionization absorption features dominate the DLA spectrum 
({\it red line}), and the high-ionization metal lines are weak, because DLAs trace large  
column densities of neutral hydrogen gas in galaxies, typically self-shielded against 
the external radiation field ($N_{\rm H I}>2 \times 10^{20}\,{\rm cm^{-2}}$; 
see the review by \citealt{Wolfe2005}). The largest difference between the two spectra 
occurs for the N{\sc v} absorption doublet, followed by that of O{\sc vi}, which 
may illustrate the impact of turbulent and/or collisional processes in the outflow. The 
Si{\sc iv} absorption feature is weaker in the outflow than in the DLA spectrum, contrary to 
the other high-ionization doublets, which could be linked to the 
photoionization properties of the media inhabited by this species. We discuss these 
measurements in the context of the outflow structure further in \S~\ref{sec:bump}. 

\begin{figure*}\center %------------  MULTI STACK	 
\includegraphics[width=1\textwidth]{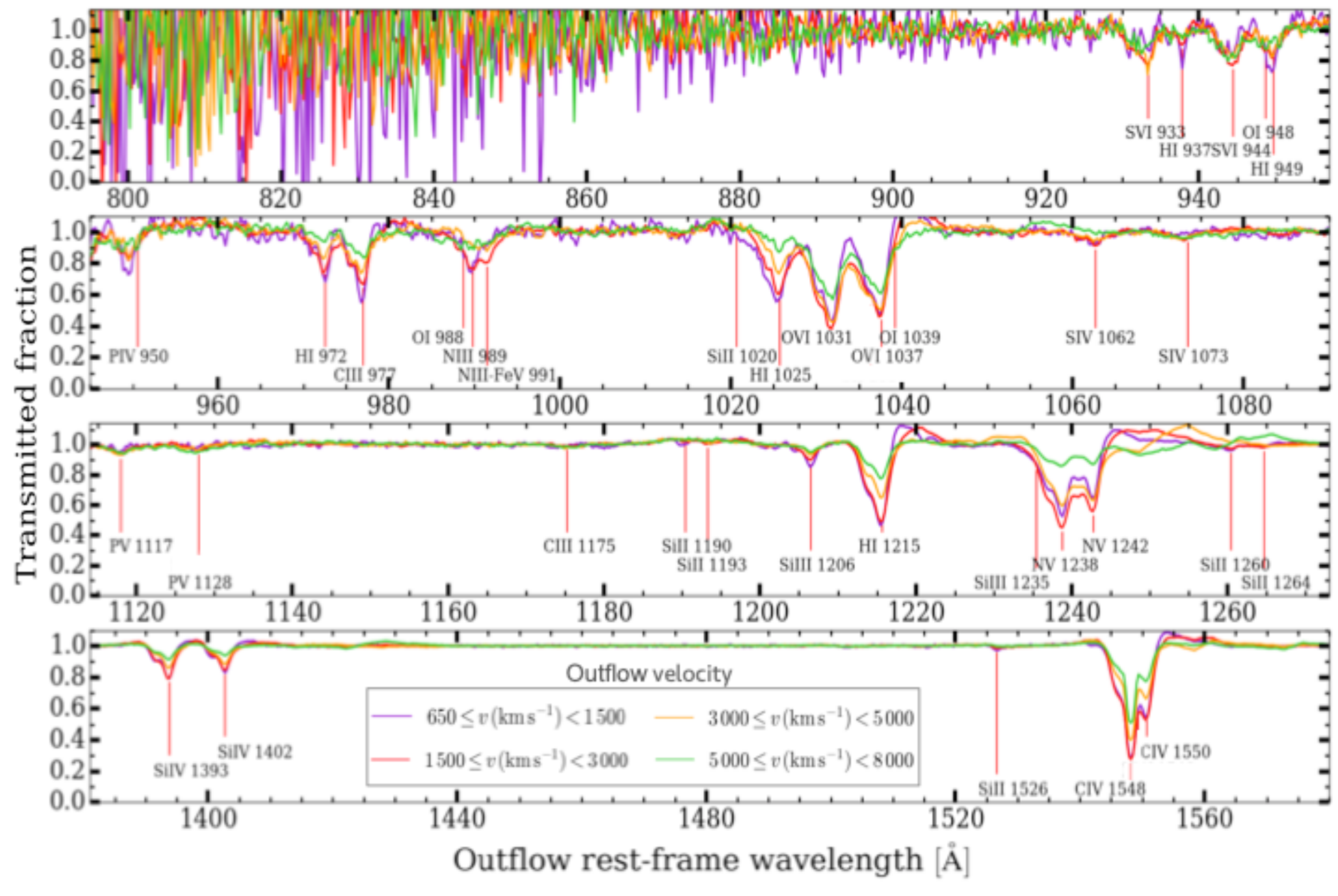}
\caption{Comparison between renormalized partial composite outflow spectra from four outflow velocity 
ranges, where we have labeled relevant transitions. Note the line-locking feature on the low-wavelength 
side of the strongest features (e.g., C{\sc iv}, Si{\sc iv}, or Lyman alpha). The species O{\sc vi} and N{\sc v} 
show the strongest features at velocities above 
$1\,500\,{\rm km\,s^{-1}}$ ({\it red} and {\it orange lines}), while for ions such as C{\sc iii}, Si{\sc iii} 
and the H{\sc i} transitions other than Lyman alpha, this happens for the outflow velocity range 
between $650 - 1\,500\,{\rm km\,s^{-1}}$ ({\it violet line}). Furthermore, the doublets C{\sc iv}, Si{\sc iv}, 
and the Lyman-alpha line show the same strength in the {\it red} and {\it violet lines}. These observations   
suggest different phases and/or different physical processes driving these ions (see main text). The 
composites reveal significant mean outflow absorption features within the Lyman-alpha forest down to 
$\approx 920 \angs$, although strong lines are observed at shorter wavelengths in individual cases.}
\label{fig:multi}
\end{figure*}

    Figure \ref{fig:multi} displays a comparison of outflow composites for four different 
ranges of outflow velocities, and at wavelengths below $1\,680 \angs$  where most of the 
outflow UV absorption occurs. We have renormalized the spectra for visualization, and 
labeled the transitions most relevant for our discussion, with the line-locking feature visible 
on the low-wavelength side of the strongest ones (e.g., C{\sc iv}, Si{\sc iv}, or Lyman alpha). 
Despite the noise, Figure \ref{fig:multi} suggests that the strongest features of the doublets  
O{\sc vi} and N{\sc v} appear at outflow velocities above $1\,500\,{\rm km\,s^{-1}}$ ({\it red} 
and {\it orange lines}). However, for ions such as C{\sc iii}, Si{\sc iii} 
and the H{\sc i} transitions other than Lyman alpha, this seems to happen for the outflow velocity range 
between $650 - 1\,500\,{\rm km\,s^{-1}}$ ({\it violet line}). Also, the doublets C{\sc iv}, Si{\sc iv}, 
and the Lyman-alpha line show the same strength in the {\it red} and {\it violet lines}, contrary 
to the previous ions. In the next section, we will show that the quantification of these observed 
differences suggests different phases and/or different physical processes driving these features 
in the outflow. 

      Figure \ref{fig:multi} also shows that the average impact of the outflow absorption features 
in the Lyman-alpha forest is important down to wavelengths of $\approx 920 \angs$. This 
effect arises mostly from the Lyman transitions of hydrogen, high-ionization species such 
as P{\sc v}, O{\sc vi}, and S{\sc vi}, and intermediate ions such as N{\sc iii} and C{\sc iii}. 
At shorter wavelengths the noise is large and the detection of lines is difficult but in individual 
cases additional features of S{\sc iv} and a few other intermediate to high ions may be present.

      We conclude by noting that we observe a large number of high- and low-ionization 
absorption features in our outflow spectra, but we find no evidence for the presence of molecular 
${\rm H_2}$ in any case. However, the line-locking components on the left-hand side of the 
absorption lines may overlap with some of the molecular hydrogen features and thus  
prevent their detection.

%--------------------------------------   DEPENDENCES -----------------
\section{Outflow Dissection}\label{sec:dependencies}

         In the next sections, we assess the dependences of the line and line-locking 
equivalent widths on outflow and quasar properties. We will refer to the respective 
absorption components as line and line-locking from now on for simplicity.

%-------------------------------------- VELOCITY  DEPENDENCE-----------------
\subsection{Outflow Velocity Dependence}\label{sec:velocity}

      We address here the evolution of the  equivalent widths with outflow 
velocity, and examine the differences between ions of different species and ionization states.

\begin{figure*}\center %------------  VELOCITY
\includegraphics[width=1.\textwidth]{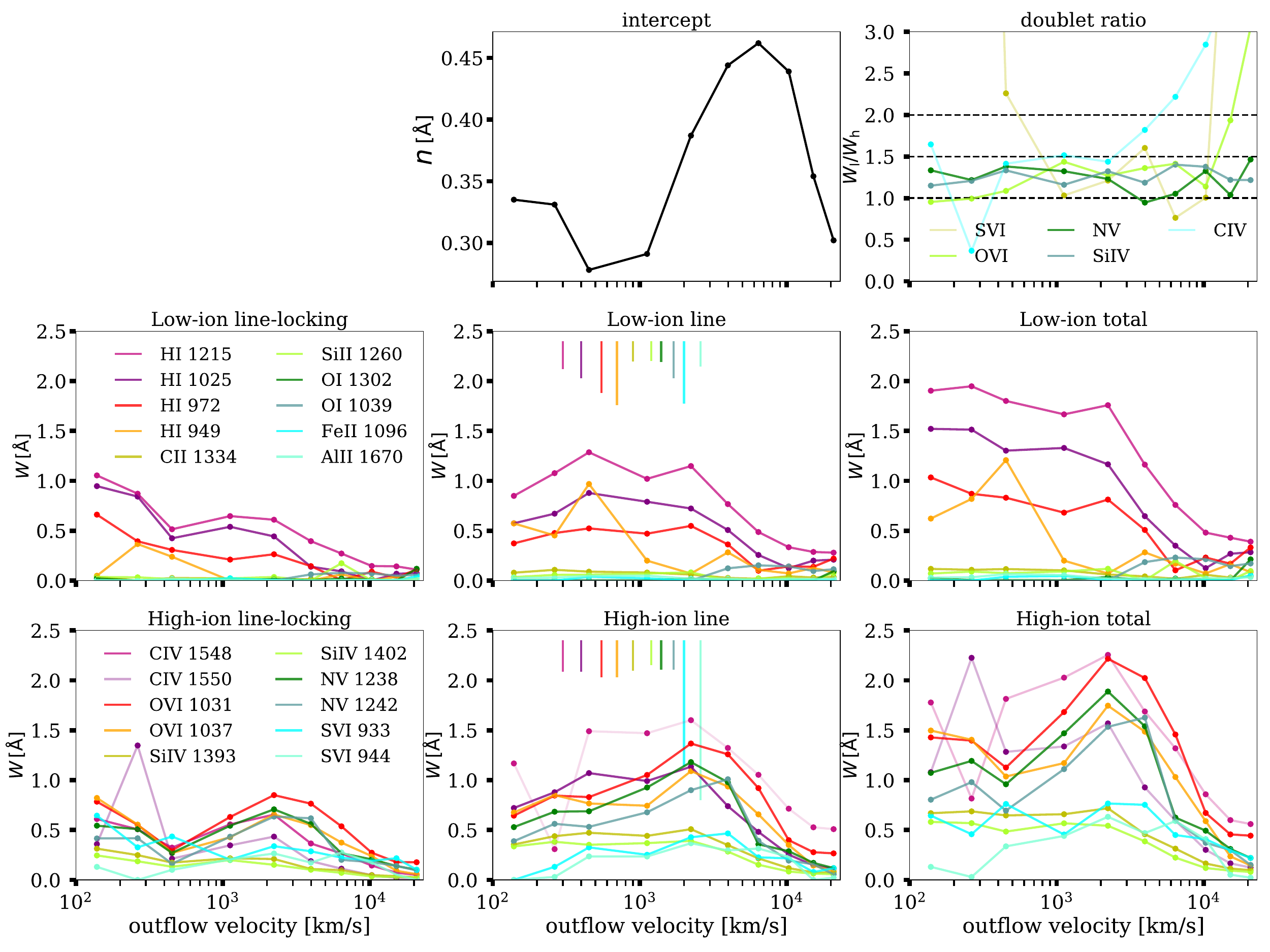}
\caption{Evolution of the equivalent width with outflow velocity. The {\it left panel} in 
the {\it top row} displays the evolution of the intercept $n$, related to the width of the 
absorption features (Eqs.~\ref{eq:fit} and \ref{eq:ap}). The ratios between the total equivalent widths of 
the two features of a doublet are shown in the {\it top right panel}, where the {\it dashed horizontal 
lines} highlight the ratio values 2, 1.5, and 1. 
The panels in the {\it middle row} display the evolution of the equivalent widths for the strongest 
low-ionization species. Those of the high-ionization species are shown in the {\it bottom row} ones.
The measured equivalent widths are represented by {\it colored points} connected by straight lines 
for visualization, and their horizontal positions correspond to the values in the second column 
of  Table \ref{ta:subsample}. The {\it leftmost panels}  in the 
two lower rows represent the values measured for the line-lockings, and the 
{\it central panels} show the values measured for the lines. The {\it rightmost panels} display 
the total equivalent width for each species, simply adding the values of the two respective panels 
on their left. The uncertainties (one sigma values) of the line equivalent widths are indicated by the  
{\it vertical lines} in the central panels following the same color code as the corresponding lines 
and line-lockings. This figure shows a quasi-constant equivalent width evolution with velocity 
below $\sim 2\,500\,{\rm km\,s^{-1}}$ for both low- and high-ionization species in general, followed by 
a decrease at higher velocities. However, a different behavior is visible for the equivalent widths of 
the O{\sc vi} and N{\sc v} transitions, which increase steeply with velocity before the overall decrease, 
thus resembling the evolution of the intercept $n$  
(see the text for more discussions on differences between species/transitions). The values of the doublet 
ratios show a combination of highly and mildly saturated lines, with a mean value at around $1.25$.}
\label{fig:velocity}
\end{figure*}

     The panels in the {\it middle row} of Figure \ref{fig:velocity} display the evolution of the 
equivalent widths for the strongest low-ionization species, and those of the high-ionization 
species are shown in the {\it bottom row}. 
The measured equivalent widths are represented by {\it colored points} connected by 
straight lines for visualization, and their horizontal positions correspond to the values 
in the second column of Table \ref{ta:subsample}. The {\it leftmost panels} in the two lower rows of the figure 
represent the values measured for the line-lockings, and the {\it central panels} show 
the values measured for the lines. The {\it rightmost panels} display the total equivalent 
width for each species, simply adding the values of the two respective panels on their left. 
For guidance, we plot the uncertainties (one sigma values) of the line equivalent widths as 
{\it vertical lines} in the central panels, following the same color code as the corresponding 
lines and line-lockings. We emphasize again that these uncertainties are derived from the 
spectral pixels and do not include the possible effect of systematics or modeling 
(see \S~\ref{sec:eqwerr}). The left panel in the {\it top row} displays the evolution of the 
intercept $n$, related to the width of the absorption features through Eqs.~\ref{eq:ap} and 
\ref{eq:fit}, and the right panel represents the total equivalent width ratios between the two 
transitions of atomic doublets (short over long wavelength). 
Here, the {\it dashed horizontal lines} highlight the ratio values 
2, 1.5, and 1, which describe an increasing saturation level, from non-saturated to completely 
saturated absorption features. The format just described for this figure will also be used when 
addressing the dependences of the equivalent widths on other parameters, unless stated otherwise.

\begin{figure*}\center %------------  derivatives
\includegraphics[width=0.5\textwidth]{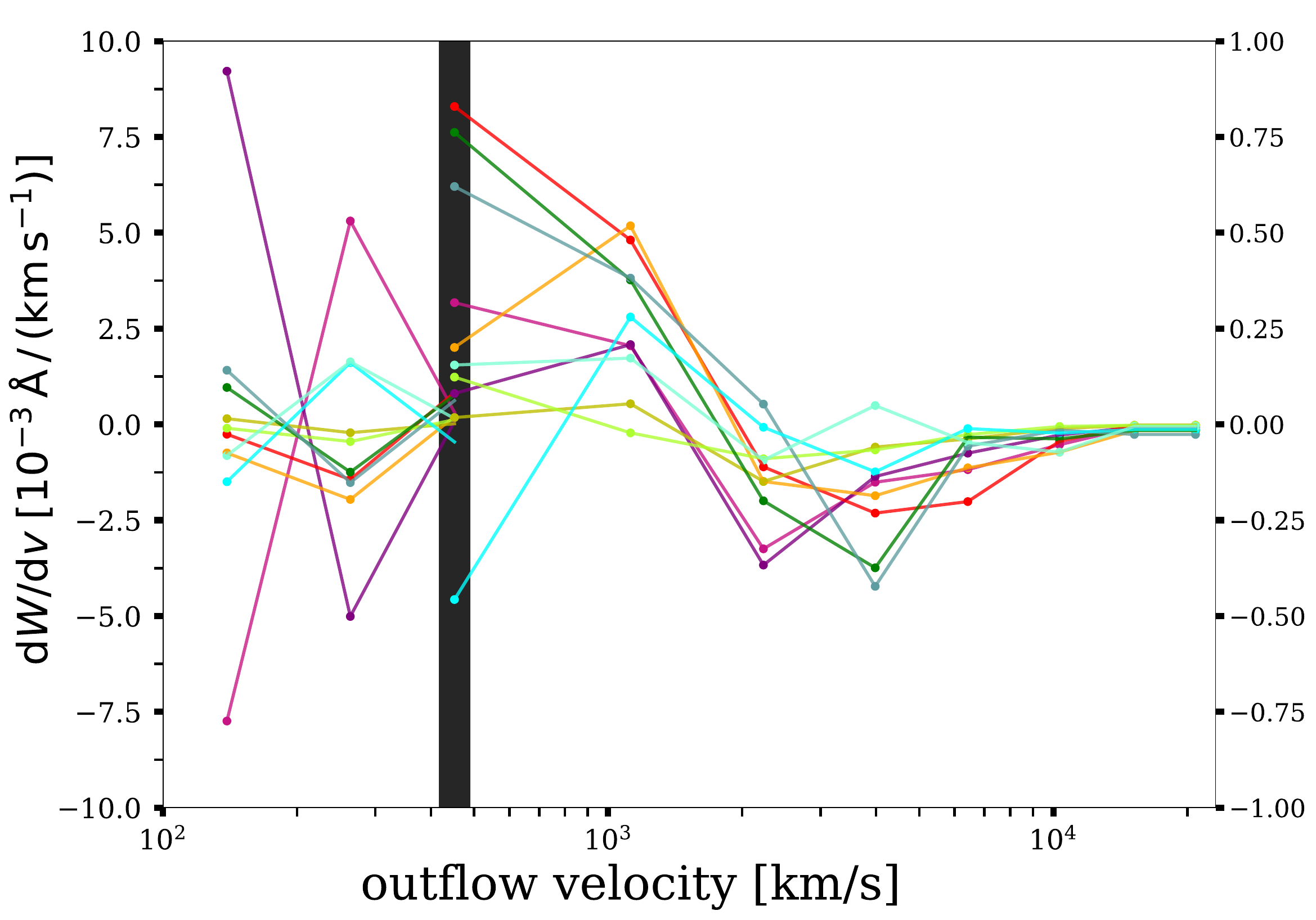}\includegraphics[width=0.47\textwidth]{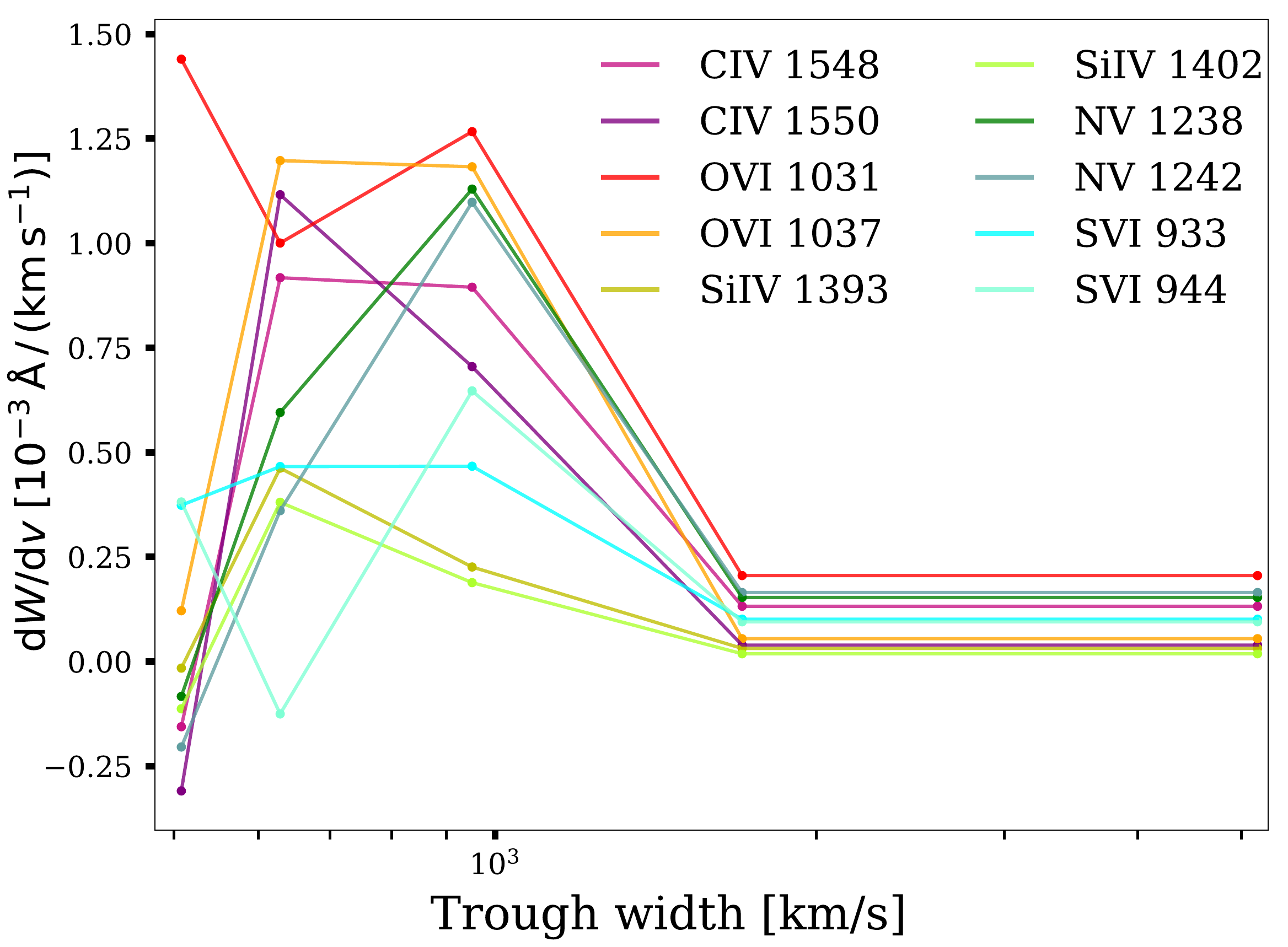}
\caption{{\it Left panel}: evolution of the derivative of the equivalent width as a function of outflow 
velocity. The panel is divided by the {\it black vertical band} in two regions with different scales. 
Within the velocity range $500-1\,000\,{\rm km\,s^{-1}}$ the highest values of the derivative correspond 
to the O{\sc vi} and N{\sc v} lines, indicating their steepest evolution compared to the other ions in that 
region. {\it Right panel}: same as left panel but for trough width. The highest values below the first 
$\sim 2\,000\,{\rm km\,s^{-1}}$ generally correspond to the O{\sc vi} and N{\sc v} lines, closely followed 
by those of C{\sc iv}. The lines of S{\sc vi} show lower values but generally still above those of Si{\sc iv}. 
We have omitted the uncertainties in this calculation for simplicity 
and visualization, but we expect the differences in the values to be of the order of the uncertainties 
above $\sim 1\,000\,{\rm km\,s^{-1}}$.}
\label{fig:dervel}
\end{figure*}

     Figure \ref{fig:velocity} shows a quasi-constant equivalent width for the low-ionization 
species in the first $\sim 2\,500$ ${\rm km\,s^{-1}}$ in general ({\it middle row}), although differences 
between the line and line-locking at the smallest velocity bins are visible. This is because 
the data points are located  at the velocity position of the troughs in all cases, although the 
line-locking denotes velocities $\approx 500$ ${\rm km\,s^{-1}}$ larger than those of the lines. 
The equivalent widths of the low-ionization lines ({\it panel in the 
center}) are smaller in the first $\sim 500$ ${\rm km\,s^{-1}}$ than at larger velocities. This may reflect 
the impact to the absorption features of the gas in the quasar host 
halo, not in the outflow. Halo gas is less metal enriched and moves at the typical virial velocities of 
$200-300\,{\rm km\,s^{-1}}$ for halos of mass $M_{\rm h}$ a few times $10^{12}\,{\rm M_{\odot}}$ 
at our average redshift of $z\sim 2.4$. 
Furthermore, stellar winds with velocities of several hundreds of 
${\rm km\,s^{-1}}$ could also contribute to this velocity range. The equivalent widths of both low- and 
high-ionization species show a decrease above $2\,500 - 4\,000\,{\rm km\,s^{-1}}$.  This decrease 
may denote a transition between two media with different physical properties, e.g., metal abundance 
or density. This could happen, for instance, in the transition region between the circumnuclear parts 
close to the black hole and the less dense interstellar medium, but this needs to be tested by future 
analysis.  
   
     The general evolution of the equivalent width for the high ions is considerably flat before the 
steep decrease at the same velocity as the low ions. However, a different behavior is visible for 
the transitions of the O{\sc vi} and N{\sc v} ions, which resembles the shape of the evolution of   
the intercept, $n$ (this is most visible comparing the evolutions for the intercept and those in the 
{\it bottom right} panel of Figure \ref{fig:velocity}). The equivalent widths of these two species 
increase steeply with velocity after $\sim 500\,{\rm km\,s^{-1}}$, before peaking and decaying 
quickly after  $\sim 2\,500\,{\rm km\,s^{-1}}$. This arises from the fact that while all the absorption 
features in general broaden with velocity following the evolution of $n$, their depth decreases, except 
for the cases of O{\sc vi} and N{\sc v} as illustrated by Figure \ref{fig:multi}. 
The left panel in Figure \ref{fig:dervel} shows the derivative of the equivalent widths with respect 
to outflow velocity to emphasize the different trends of the high-ionization species. This panel is 
divided in two parts  with different scales by a {\it thick vertical line}. The left part covers a broader 
range to include the large values of C{\sc iv} while the right part zooms-in by a factor of ten. As already 
suggested in Figure \ref{fig:velocity}, between the velocity range $\sim 500-1\,000\,{\rm km\,s^{-1}}$ the 
highest values of the derivatives correspond to the O{\sc vi} and N{\sc v} lines, indicating their steeper  
evolution compared to the other ions. We have omitted the uncertainties in this calculation for simplicity 
and visualization, but above $\sim 1\,000\,{\rm km\,s^{-1}}$ one expects the differences between the 
values to be of the order of the uncertainties. 
We discuss the origin of the O{\sc vi} and N{\sc v} behavior in \S~\ref{sec:bump}. 

    The total equivalent width ratios of the doublets in the {\it top right panel} indicate that the  
absorption features of these doublets are a 
combination of highly and mildly saturated transitions, with average ratio values around 
$1.25$ when the equivalent widths are measured with precision. 

    We observe a similar behavior for the equivalent widths of the line and line-locking 
absorption components in general. This suggests that the radiative effects, i.e., line locking, remain 
roughly constant and do not vary greatly with outflow velocity.

    Finally, the evolutions plotted in Figure \ref{fig:velocity} also enable us to easily identify the impact 
of degeneracies between overlapping transitions: the magenta and purple lines in the {\it bottom 
right panel} show the effect of the overlapping line-locking C{\sc iv}$\,\lambda1550$ and line 
C{\sc iv}$\,\lambda1548$ features in the second lowest velocity bin, respectively. 
Overlapping and noise effects on the doublets can also be inferred from the {\it top right 
panel}, since the measurements of the ratios are highly sensitive to such contaminants that 
shift their values away from the theoretical range within the values 1 and 2.

\begin{figure*}\center %------------  WIDTH
\includegraphics[width=1.\textwidth]{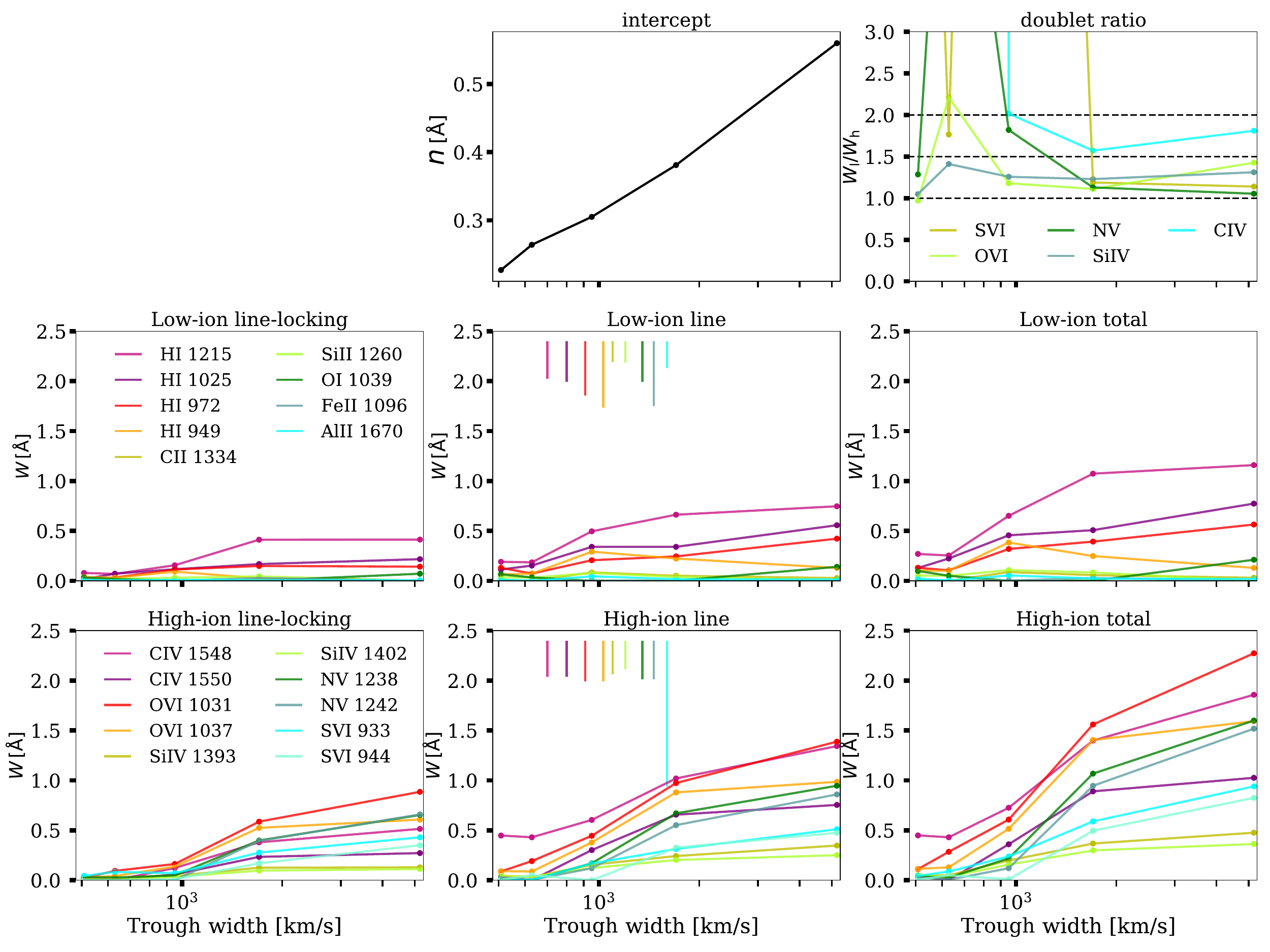}
\caption{Evolution of the equivalent width as a function of trough width, 
following the format of Figure \ref{fig:velocity}. A general increase of 
the equivalent width with velocity is visible, although this is steeper for  
the high-ionization species than for the low-ionization ones.}
\label{fig:width}
\end{figure*}

\begin{figure*}\center %------------  VELO-WIDTH
\includegraphics[width=1.\textwidth]{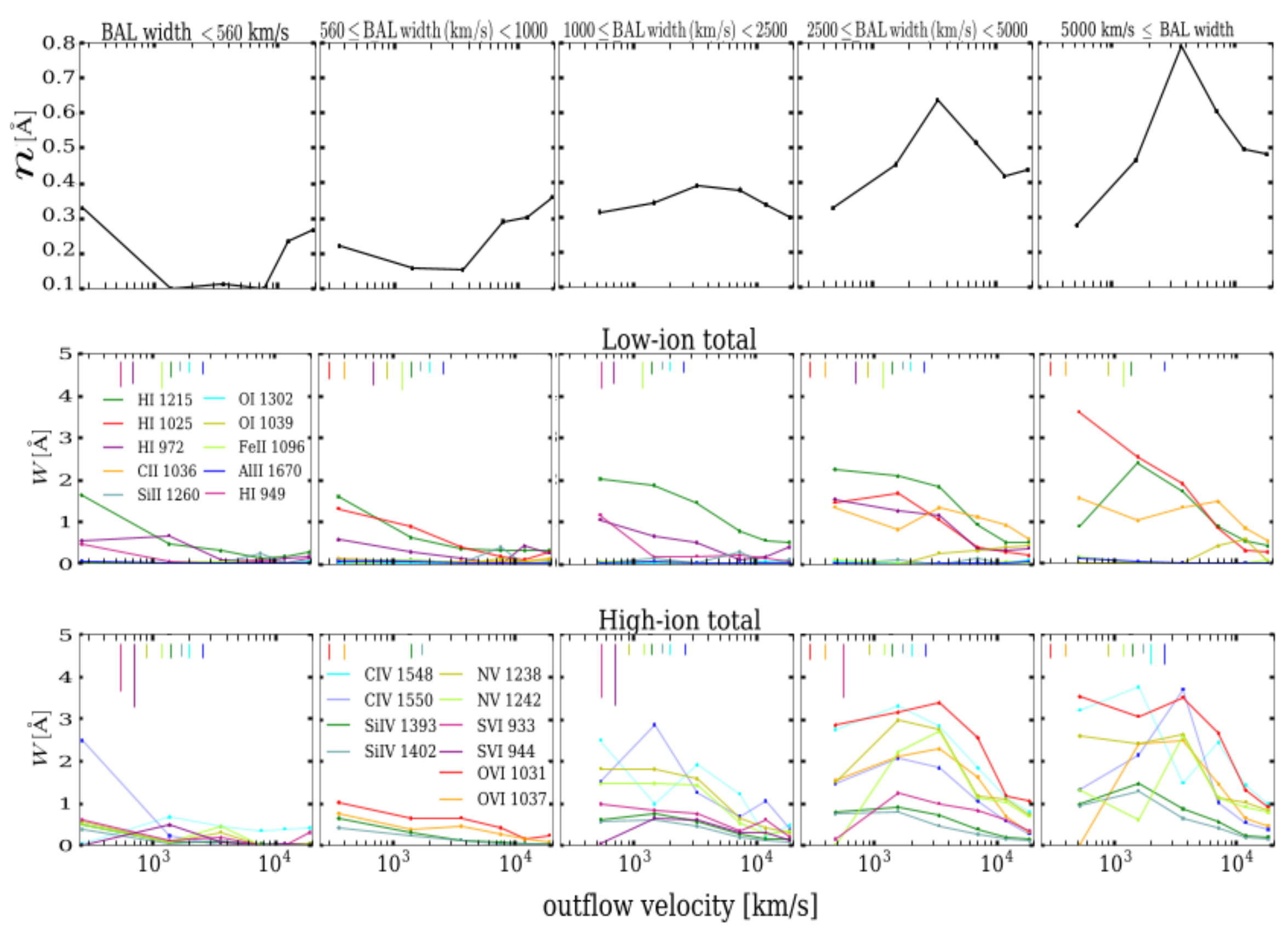}
\caption{Evolution of the total equivalent widths of low-ionization ({\it middle panels}) and 
high-ionization ({\it bottom panels}) species with outflow velocity and trough width. Every {\it column} 
denotes a trough width range, increasing toward the right. Each {\it top panel} represents the 
measurements of the intercept (related to the line width) with outflow velocity, where the values 
in the range $\sim 2\,000 - 6\,000\,{\rm km\,s^{-1}}$ show a significant increase with trough width, 
while the values in the laterals remain almost constant. The equivalent widths of the low-ionization 
species increase with trough width in general, with a negative-slope dependence on velocity. The 
high-ionization species follow a similar behavior but the differences of O{\sc vi} and N{\sc v} 
compared to the other ions discussed in the velocity case are larger when increasing the trough widths.}
\label{fig:big}
\end{figure*}

%-------------------------------------- WIDTH ---------------------------------------
\subsection{Trough Width Dependence}\label{sec:width}

     Figure \ref{fig:width} illustrates the dependence of the equivalent widths on the width of the 
absorption troughs, following the same format as in Figure \ref{fig:velocity}. A clear increase of 
the equivalent widths with trough width is visible, but the slope is steeper for the high-ionization 
species in general, and especially for the O{\sc vi} and N{\sc v} transitions,  supported by the 
values of the derivative of the equivalent widths in the {\it right panel} of Figure \ref{fig:dervel}. 
The equivalent width ratio values for the doublets are within the range 
$1 - 1.5$, similar to those in Figure \ref{fig:velocity}.

     Below we test whether the width of the troughs has an impact on the outflow velocity dependence 
observed in the previous section.

%-------------------------------------- VELOCITY AND WIDTH ---------------------------------------
\subsubsection{Joint Velocity and Width Dependence}\label{sec:velocitywidth}
 
        We explore here the joint dependence of the equivalent widths on outflow velocity and trough width. 
This is illustrated in Figure \ref{fig:big}, where 
every {\it column} denotes a range in trough width, increasing toward the right 
columns. The {\it upper panels} show the evolution of the intercept $n$, the {\it middle panels} 
the evolution for the equivalent widths of the low-ionization species, and those for the 
high-ionization ones are plotted in the {\it bottom panels}. 

      The values of the intercept in the range of outflow velocities within $\sim 2\,000 - 6\,000\,{\rm 
km\,s^{-1}}$ present a significant increase with trough width, of up to a factor of eight for the velocity bin 
centered at $\approx 3\,700\,{\rm km\,s^{-1}}$. This trend is best matched by 
the high values of the O{\sc vi} and N{\sc v} equivalent widths at the same velocity 
position, compared to a smaller increase for the other ions. 
The values of the intercept in the minimum and maximum outflow velocity points 
show little evolution with trough width, indicating that the width of the lines does not change 
significantly. These differences are consistent with our interpretation of three outflow velocity 
regions representing different gas phases or spatial locations discussed in \S~\ref{sec:velocity},    
as well as also argued by \cite{Nestor2008}. 

      In conclusion, the differences between the equivalent widths of O{\sc vi} and N{\sc v} 
compared to those of other ions are maximal at outflow velocities of a few thousands  
of ${\rm km\,s^{-1}}$ and for troughs broader than $\sim 2\,500$ ${\rm km\,s^{-1}}$. 

     Figure \ref{fig:big} also shows that the equivalent widths of the high-ionization species 
are generally significantly larger in the two rightmost columns than in the others, and specifically at 
outflow velocities above $1\,000 - 2\,000\,{\rm km\,s^{-1}}$.  These width and outflow velocity 
values broadly match the strict definition of BALs (i.e., outflow velocities $\gtrsim 3\,000\,{\rm km\,s^{-1}}$ 
and trough widths $\gtrsim 2\,000\,{\rm km\,s^{-1}}$). The physical process/es that boost the equivalent 
width values of some high-ionization species (see our discussion in \S~\ref{sec:collis}) in the two 
rightmost panels, seem to not be present in the other three leftmost columns. This effect may be 
the driver of the differences observed between BALs and mini-BALs/NALs, although we are here 
analyzing absorption lines while the outflow-type names arise from the observation of the broad 
troughs in the quasar spectra. 

%-------------------------------------- MIN VELOCITY ---------------------------------------
\subsection{Trough Minimum Velocity Dependence}\label{sec:minvel}

       We assess now the dependence of the  equivalent widths on the degree of 
detachment of the troughs, motivated by polarization results that suggest that 
the polarization of the outflow decreases with the velocity distance from the source 
\citep[e.g.,][]{Lamy2004}.

    We show in Figure \ref{fig:vmin} in the Appendix the dependence of the equivalent widths on 
the minimum velocity of the trough. The evolutions match those found with respect to the 
outflow velocity within the uncertainties, and additional effects arising from the  detachment 
are not observed. This is not surprising because our trough widths are generally narrow, with their 
distribution peaking at $\sim 700\,{\rm km\,s^{-1}}$ (Figure 1 in \citealt{Masribas2019}), and 
so the minimum trough velocities are similar to the velocities of the outflow. 

     We discuss the relation between the 
degree of polarization and the velocity of the outflows in more detail in \S~\ref{sec:polarization}.

%-------------------------------------- QSO MAGNITUDE-----------------
\subsection{Quasar Magnitude and Redshift Dependences}\label{sec:magnitude}

        Figure \ref{fig:mag} in the Appendix shows the dependence of the equivalent 
   widths on the absolute magnitude 
of the host quasars, where no evolution is detected. For the case of the lines, some high-ionization 
species ({\it middle bottom panel}) tentatively suggest a decrease of the equivalent width with 
increasing quasar brightness, consistent with the trend of the intercept ({\it top left panel}). However, 
this decrease is neither visible for the line-locking nor the low-ionization features, and the variations are at 
the level of the uncertainties in the measurements. We discuss the lack of 
dependence on the quasar magnitude and compare it with other works in \S~\ref{sec:eqwmag}.  

      Figure \ref{fig:redshift} in the Appendinx illustrates the dependence of the equivalent widths 
on quasar redshift. 
The measurements remain constant over the whole redshift range, indicating that the 
equivalent width values are mostly driven by local processes, and are independent of the cosmic 
metallicity evolution. It is worth noting, however, that our measurements cover a narrow redshift 
range compared to the overall period of cosmic metal enrichment.

%-------------------------------------- OUTFLOW STRUCTURE ---------------------------------
\section{The Multiphase Outflow Structure}\label{sec:bump}

         Figures \ref{fig:multi} and  \ref{fig:velocity}, together with Figure \ref{fig:dervel}, showed that the 
dependence on outflow velocity of the O{\sc vi} and N{\sc v} transitions is significantly different from 
that seen for other low- and high-ionization species. 
The behavior of these two ions resembles that of the intercept ($n$; {\it top 
left panel} in Figure \ref{fig:velocity}), the latter related to the width of the absorption features (see 
also the evolution of the intercept and these two species in Figure \ref{fig:big}, and a 
quantitative estimate of this relation through a Spearman rank correlation analysis in \S~\ref{sec:spear} 
in the appendix). Similar 
differences are also suggested in the bottom right panel of Figure \ref{fig:width}, where the 
features of O{\sc vi} and N{\sc v} show a steeper slope compared to the other high-ionization 
species in the trough width range $\sim 1\,000 - 2\,000\,{\rm km\,s^{-1}}$, and in some cases also 
at larger width values. Furthermore, the bottom right panel in Figure \ref{fig:width} tentatively 
suggests that the slope of the equivalent width evolution with trough width might be steeper for 
C{\sc iv} and S{\sc vi} than for Si{\sc iv}, although flatter than those of O{\sc vi} and N{\sc v}. 
This might also be the scenario illustrated in the bottom middle panel of Figure \ref{fig:velocity}, 
 supported by the values of the derivatives of the equivalent widths with trough width in the right 
panel of Figure \ref{fig:dervel}.  
The uncertainties in these last cases, however, are large and the suggested trends may not be real.

%-------------------------------------- russiand doll ---------------------------------
\subsection{The Russian-doll Model}\label{sec:doll}

       We propose a multiphase stratified structure for the outflow  in order to explain the different 
evolutions of the species, inspired by that proposed by \cite{Stern2016} for 
the circumgalactic medium. In the typical conditions of the 
circumgalactic medium, the O{\sc vi} and N{\sc v} species trace gas regions that 
are less dense and slightly hotter ($T\sim 2-3\times 10^5$ K, $n_{\rm H}\sim 5\times 
10^{-5}\,{\rm cm^{-3}}$) than those inhabited by C{\sc iv} and S{\sc vi} ($T\sim 1-2\times 
10^5$ K, $n_{\rm H}\sim 2\times 10^{-4}\,{\rm cm^{-3}}$), and especially by Si{\sc iv} 
($T\sim 7\times 10^4$ K, $n_{\rm H}\sim  10^{-3}\,{\rm cm^{-3}}$) on average 
\citep[see figure 6 in][and figure 1 in \citealt{Stern2016}]{Tumlinson2017}.

\begin{figure}\center %------------  FLOW STRUCTURE
\includegraphics[width=0.45\textwidth]{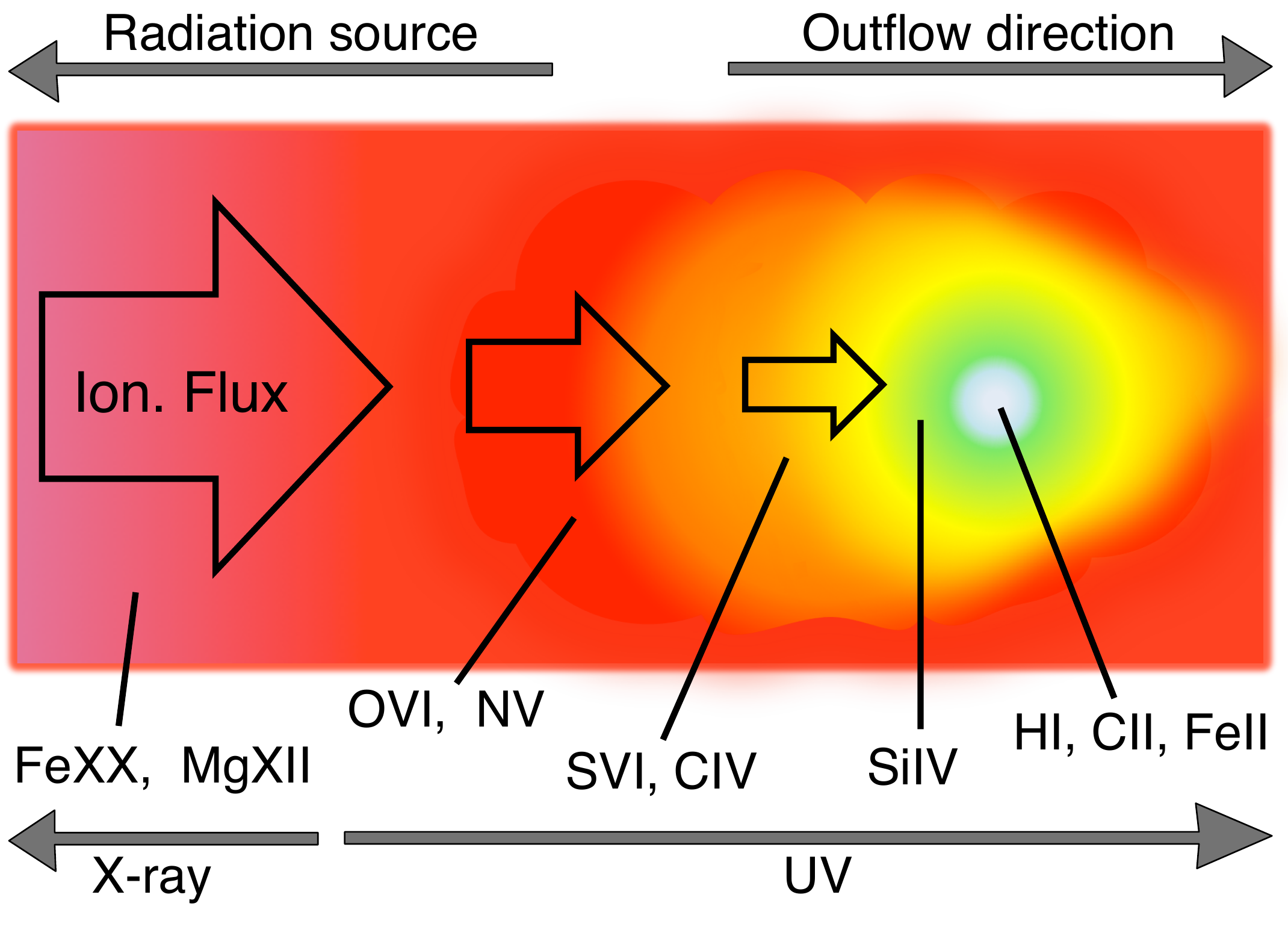}
\caption{Cartoon of the proposed multiphase structure of the gas in outflows. From the outer 
to the inner regions, the ionizing radiation indicated by the {\it arrows} is attenuated and the gas 
has a lower temperature and higher density, thus being able to host low-ionization species in the 
core.}
\label{fig:flow}
\end{figure}

     Figure \ref{fig:flow} illustrates a cartoon of our proposed Russian-doll model for the multiphase 
outflowing gas.  Here, O{\sc vi} and N{\sc v} trace the outermost UV 
parts of the outflowing material ({\it red area}), exposed to the strong radiation field ({\it large 
arrow}) that keeps them highly ionized. A second inner region ({\it orange-yellow area}) may then 
contain the C{\sc iv} and S{\sc vi}, where the ionizing flux has already been attenuated considerably 
({\it middle arrow}) by the first UV layer, allowing a higher density and a lower temperature. Two 
additional regions deeper in the gas ({\it green} and {\it light blue regions}), at lower temperatures, 
higher densities, and less exposed to the radiation ({\it small arrow}), host the Si{\sc iv} and the 
low-ionization species, respectively. A Si{\sc iv} protected from the strong radiation field 
in the deepest regions of the outflowing material can explain the similarity between the equivalent 
width evolution of Si{\sc iv} and the low-ionization species, and the fact that Si{\sc iv} is the only high-ionization 
species that shows a weaker feature in quasar outflows than in DLAs in our Figure \ref{fig:stack}. 
However, we saw in \cite{Masribas2016c} that the equivalent width of Si{\sc iv} does not 
correlate with the column density of neutral hydrogen in DLAs (lower panels in figure 11 there), 
which suggests different gas phases for Si{\sc iv} and the low ions. Intermediate-ionization species 
such as Al{\sc iii} likely reside in the two inner regions as well, as they often show a similar behavior 
to that of low ions \citep[e.g.,][]{Filizak2014}.

%-------------------------------------- Radiation Shield ---------------------------------
\subsection{The Radiation Shield}\label{sec:shield}

         Our outflow structure could also account for 
the `radiation shield' that is often invoked in radiatively-accelerated outflow scenarios 
\citep[][although see \citealt{Baskin2013,Baskin2014b}]{Murray1995,Chelouche2003}. In brief, 
most of the acceleration arises from the absorption of UV radiation, but if the gas is highly-ionized it  
is too optically thin to the UV photons and the radiative acceleration becomes inefficient 
\citep[the so-called overionization problem;][]{Castor1975,Stevens1990,Higginbottom2014,
Dannen2018}. Therefore, the high-energy (X-ray and far-UV) photon flux 
needs to be significantly suppressed before reaching the acceleration region to keep the medium 
optically thick to UV radiation \citep{Leighly2004,Baskin2013}, which can be 
achieved with a large column density, above $N_{\rm H} \sim 10^{22} - 10^{23} \,{\rm cm^{-2}}$, of 
highly-ionized material \citep{Murray1995}. This shield  is sometimes assumed to have a constant 
density and to remain static close to the source of radiation,  but \cite{Hamann2013} indicated 
that absorption signatures from this gas at very low velocities are not observed \citep[see an extended 
discussion in section 5.1 of][]{Hamann2018}. Additionally, \cite{Chartas2002,Chartas2003} detected 
broad X-ray absorption features at velocities $\gtrsim 0.2 - 0.3c$ that disfavor the static medium. 
We suggest that the radiation-shielding gas is comoving with the outflow, more precisely, that the radiation 
shield {\it are} the most external layers in the multiphase structure of Figure \ref{fig:flow} (leftmost {\it red} and 
{\it purple} areas), where X-ray and far-UV transitions take place \citep[][see also 
\citealt{Hamann2018b}]{Mckernan2007,Reeves2013}. \cite{Proga2000} found that the shield arises 
naturally in a disk wind scenario \citep[see also the radiative transfer simulations in a clumpy outflow 
by][]{Matthews2016}. 

          \cite{Baskin2014b} argued that a stratified multiphase structure as the one proposed 
here could circumvent the overionization problem without the need of a static shielding layer 
\citep[see also][]{Dekool1995,Arav2013}, and a structure with different phases was also  suggested by 
\cite{Baskin2014} and \cite{Stern2014} to explain observed properties of the broad line region 
(BLR) in AGNs.

%-------------------------------------- collisions ---------------------------------
\subsection{The Collisional Processes}\label{sec:collis}

      With this outflow structure in mind, we argue that the equivalent width differences observed 
between species result from, and reveal, the effect of collisional processes. Collisions would  
be most important in the external layers where the temperatures are the highest, and therefore 
more visible for the O{\sc vi} and N{\sc v} ions. The presence 
of collisions is further supported by the detection of the S{\sc iv}*$\,\lambda 1073$ and 
C{\sc iii}*$\,\lambda 1175$ absorption lines (the latter at a considerable lower significance as 
visible in Figure \ref{fig:multi}), as both arise from excited metastable states 
populated through collisional processes \citep{Leighly2009}. However, we only detect 
these transitions in the outflow  spectra where the O{\sc vi} and N{\sc v} equivalent widths 
are significantly large. This is illustrated in the upper panel of 
Figure \ref{fig:sivcoll}, where the excited-state transition S{\sc iv}*$\,\lambda 1073$ is mostly detected 
in the two subsamples with mean outflow velocities $\sim 2\,200\,{\rm km\,s^{-1}}$ ({\it red 
line}) and $\sim 4\,000\,{\rm km\,s^{-1}}$ ({\it orange line}). The lower panel in Figure 
\ref{fig:sivcoll} quantifies these observations through the equivalent width of the two sulfur lines, 
compared to the scaled values of the other high ions with outflow velocity (the {\it gray lines} are the 
same as in the bottom-right panel of Figure \ref{fig:velocity}, reduced by a factor of five). Because 
these lines are very weak and our fitting algorithm does not perform reliably in this case, we compute 
the equivalent widths by summing the flux in the composite spectrum (with positive sign for flux 
below the unity and negative above). For this measurement we use the same windows as those for the 
equivalent width uncertainty calculations in \S~\ref{sec:eqwerr}. The uncertainty in the equivalent widths 
are also obtained as in \S~\ref{sec:eqwerr}, but in this case we extract them directly from the stacked 
spectrum instead of all the intervening individual spectra. Because our uncertainty estimation 
does not capture the effect of systematics, such as the calculation of the smooth continuum for the 
final normalization of the spectrum, we consider the uncertainties in Figure \ref{fig:sivcoll} to 
underestimate the true ones. Therefore, only the large positive equivalent width values should be 
considered for interpreting the results of Figure \ref{fig:sivcoll}.  The S{\sc iv*} and C{\sc iii*} 
features are also visible  in the two subsamples with mean trough widths of $\sim 1\,700\,{\rm 
km\,s^{-1}}$ and $\sim 5\,200\,{\rm km\,s^{-1}}$ (see Table \ref{ta:subsample}). 

       We note that collisional processes and shocks have also been recently 
inferred in the circumnuclear regions of the AGN host galaxy NGC 3393 by 
\citealt{Maksym2018} (see also references therein), from the analysis of X-ray 
and optical observations at high spatial resolution of the innermost parts of this galaxy. 

%-------------------------------------- waves ---------------------------------
\subsection{The Pressure Waves}\label{sec:waves}

      The detection of S{\sc iv}*$\,\lambda 1073$ and C{\sc iii}*$\,\lambda 1175$ implies high 
electron densities in the outflow, with values typically above $n_{\rm e}=10^3-10^4\,{\rm 
cm^{-3}}$ resulting from the S{\sc iv} lines \citep{Chamberlain2015,Xu2018}, 
and as high as $n_{\rm e}\gtrsim 10^9\,{\rm cm^{-3}}$ when C{\sc iii}* is also detected 
\citep{Bromage1985,Kriss1992}. The value of the density is derived from the column density 
ratios between the excited and ground state transitions of each species, i.e., S{\sc iv}$\,\lambda 
1062$ and C{\sc iii}$\,\lambda 977$, respectively. We defer detailed calculations of the electron 
density to future work, but perform a rough estimate of the values expected from our 
measurements following the discussion in section 2 of the recent paper by \cite{Arav2018}. 
      
\begin{figure}\center %------------  SIV Collisions
\includegraphics[width=0.45\textwidth]{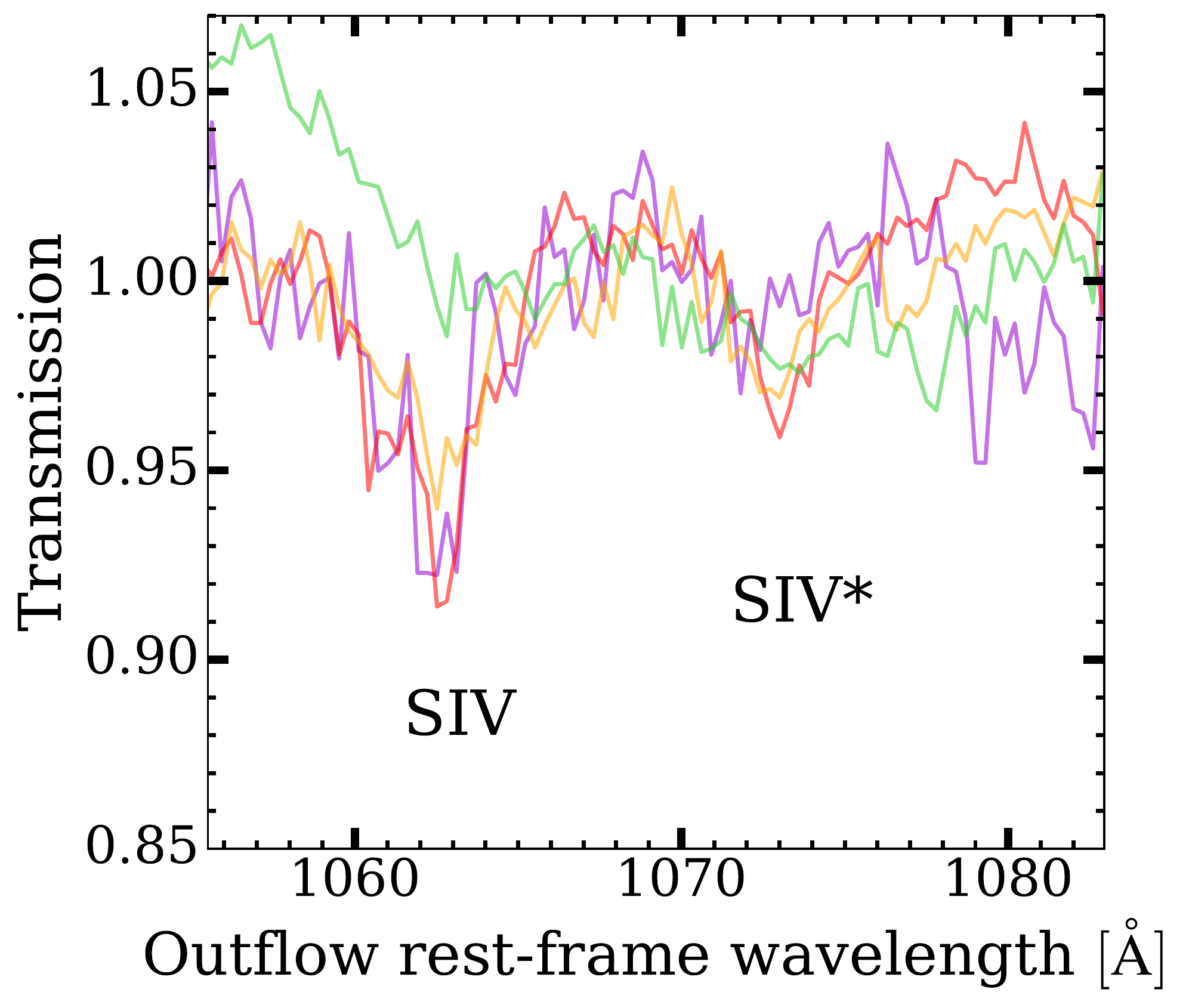}
\includegraphics[width=0.45\textwidth]{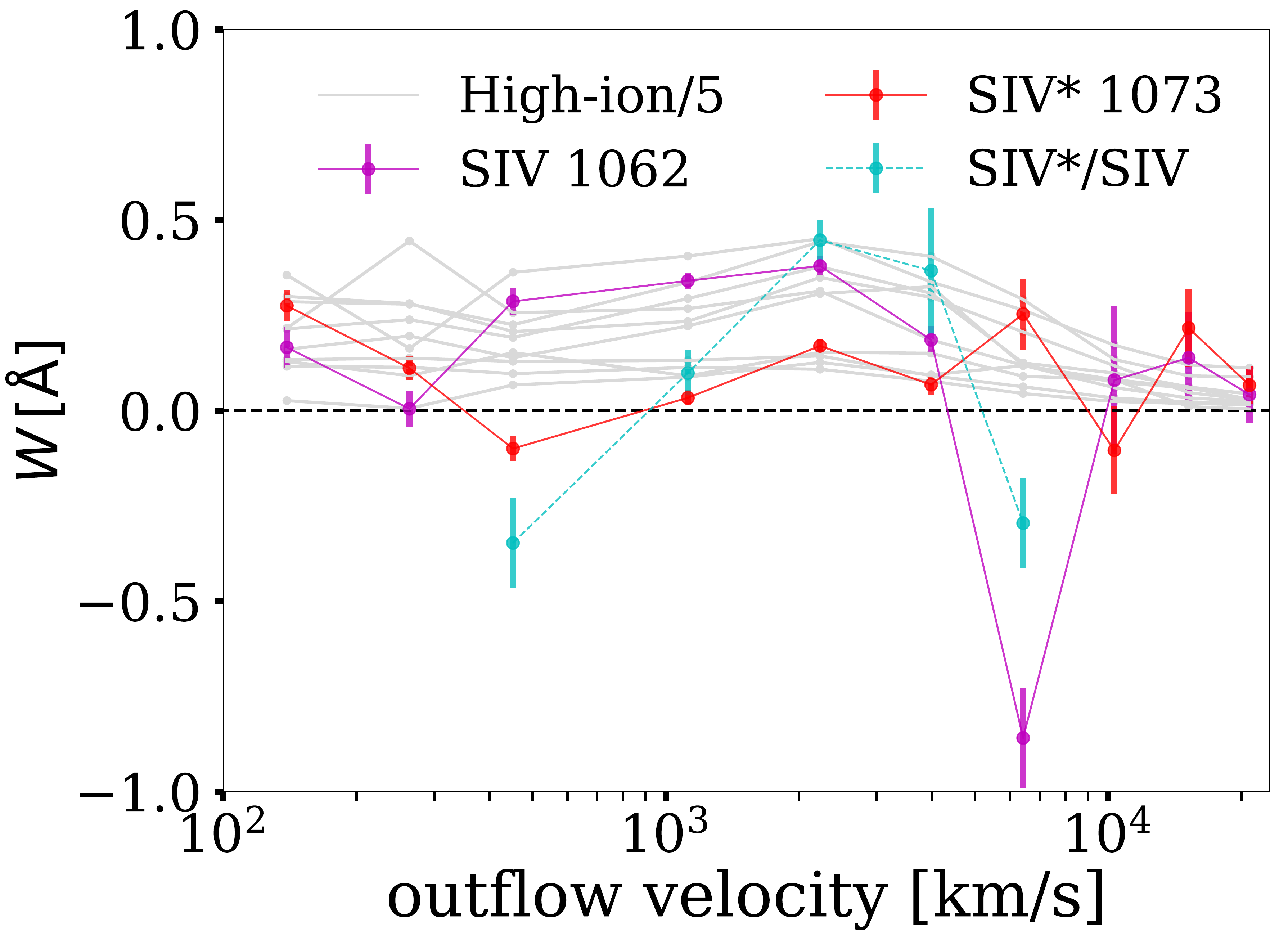}
\caption{{\it Top panel:} rest-frame outflow wavelength range illustrating the absorption features of 
the ground- and excited-state transitions S{\sc iv}$\,\lambda 1062$  and S{\sc iv}*$\,\lambda 1073$, 
respectively, for different samples of outflow velocity. The transition S{\sc iv}*$\,\lambda 1073$ is the 
strongest in the same two velocity bins ({\it red} and {\it orange lines}) where O{\sc vi} and N{\sc v} are 
also the strongest, while barely visible at lower and higher velocities. Note the 
tentative presence of the C{\sc iv} line-locking feature of S{\sc iv}$\,\lambda 1062$ at $\approx 
1\,060 \angs$ in the two strongest cases. {\it Bottom panel:} 
equivalent width with outflow velocity for the two sulfur transitions ({\it violet} and 
{\it red lines}). The {\it blue dots} and {\it dashed line} denote the values of the equivalent width ratio 
of the two transitions. In the central velocity range, where the two lines are well measured, the ratio 
increases with velocity and peaks at a value of $\approx 0.4$.  This suggests an evolution of the 
electron density correlated with that of the equivalent widths of O{\sc vi} and N{\sc v}, although 
systematic uncertainties not captured by our computations can be important.} 
\label{fig:sivcoll}
\end{figure}

       In the cases that we detect both S{\sc iv} transitions reliably 
(between $1\,000 - 5\,000\,{\rm km\,s^{-1}}$ where both lines show positive equivalent width values), 
the strength of the S{\sc iv}*$\,\lambda 1073$ 
line is smaller than that of S{\sc iv}$\,\lambda 1062$ (lower panel in Figure \ref{fig:sivcoll}), with observed 
ratios within the range $\sim 0 - 0.4$. The ratio values indicated as the {\it blue dots} and {\it dashed line} 
increase in this narrow velocity range with outflow velocity and peak at the same velocities 
where we find the  maximum equivalent widths of O{\sc vi} and N{\sc v} ({\it red} and {\it orange lines} 
in Figure \ref{fig:sivcoll}). However, the uncertainty of the point at $\sim 4\,000\,{\rm km\,s^{-1}}$ is 
large, and the negative values of the ratio at the largest and smallest velocity points are unphysical and 
likely driven by systematics, complicating a clear interpretation.  
Broadly comparing our ratio values with those illustrated in figures 1 and 2 in 
\citealt{Arav2018} (see also \citealt{Chamberlain2015}), we estimate that the electron density is 
$n_{\rm e}<0.5\times 10^4\,{\rm cm^{-3}}$ at low outflow velocities ($\lesssim 1\,000\,{\rm 
km\,s^{-1}}$), increasing up to almost $n_{\rm e}\sim 10^5\,{\rm cm^{-3}}$ at $\sim 2\,000-4\,000\,{\rm 
km\,s^{-1}}$.

    Our measurements thus suggest a correlation between the electron density when this 
can be well measured, the equivalent widths 
of the O{\sc vi} and N{\sc v} transitions, and the width of the absorption lines as traced by the 
intercept. This result is consistent with our interpretation that collisions drive the equivalent 
widths of O{\sc vi} and N{\sc v}, and likely also impact those from other species. Furthermore, the 
increase of density  with velocity before the equivalent-width turnover suggests a 
scenario where the outflow sweeps up, compresses and transports material while moving outwards, 
creating a pressure 
wave with higher densities at higher speeds (perhaps in the form of a shock wave for supersonic velocities). 
In the densest wave fronts, which may take the shape of sheets 
or filaments, the gas protected from the radiation field might cool, and thus result in the stratified Russian-doll 
structure proposed above and that hosts low ions in its core. The decrease of density at the largest velocities 
could again represent a change in the properties of the medium and/or the dissipation of the waves, but 
this needs to be tested with numerical calculations \citep[see calculations of the formation, acceleration and 
destruction of gas clumps in outflows in][as well as the shock-wave scenario proposed for 
FeLoBALs by \citealt{Claude2012b}]{Progawaters2015,Waters2016}. 

   Given the similar evolution 
observed for the equivalent widths of the line and line-locking absorption components in all cases, 
together with the omnipresence of radiative-acceleration effects that we detected in \cite{Masribas2019}, we 
postulate that radiation is the main driver of the pressure waves. Waves in quasar outflows where radiation 
pushes matter outwards are indeed expected from the theory of radiatively accelerated winds \citep[see][and references 
therein, and see also \citealt{Abbott1980,Sundqvist2018} for the case of stellar winds]{Feldmeier2002,Dyda2018b}. 
A wavy behavior  was already found in the disk wind simulations by \citealt{Proga2000} (see also \citealt{Dyda2018b}) 
and, more recently, the general relativistic magnetohydrodynamical simulations with radiation transfer by 
\cite{Davelaar2018}, although representing distances of a few gravitational radii from the black hole, also show 
the strong impact of radiation to the formation of wavy/filamentary structure in the outflows. Finally, a  
scenario in which quasar radiation accelerates material outwards and creates filamentary structure  similar to ours 
was investigated by \cite{Scoville1995}. These authors considered that the stellar winds of massive stars orbiting the 
supermassive black holes would release trails of dust that would then be accelerated outwards by the quasar radiation. 
The results by \cite{Scoville1995} were consistent with several observed outflow properties, such as the 
terminal velocities, gas column and electron densities, and the same methodology can be applied to the 
acceleration of gas instead of dust. 
Future analysis should also explore the potential contribution from other sources of pressure, 
such as thermal \citep{Chelouche2003} and magnetic energy \citep{Dekool1995,Everett2005}, or cosmic rays 
\citep[e.g.,][]{Dekool1997}. 

    Interestingly, \cite{Hamann2018} recently argued that LoBALs show the largest average column densities 
and highest velocities in the outflows. This result is consistent with our findings that larger velocities drive 
higher densities, which also relates to the presence of low-ionization species. High average 
densities in our proposed outflow structure could enhance the total amount of low-ionization 
gas, but would also increase the strength of absorption features from the high ions. This agrees 
with the high degree of reddening observed for LoBAL quasars \citep{Reichard2003,Ganguly2007,
Daix2012}, and with the measurements by  \cite{Allen2011,Reichard2003,Filizak2014,Hamann2018}, 
where LoBALs show absorption troughs of high-ionization species that are (at least) as large as those in 
HiBALs. Further potential connections between the structure of the outflows and the nature of LoBALs 
need to be addressed in future studies.

       Finally, the value of the electron density is often used to infer the distance between the absorbing 
material in the outflow and the radiation source (although see the different methods discussed in section 
7.1 of \citealt{Arav2018}). Assuming that the conditions of the typical outflow considered by \cite{Arav2018} 
are similar to our measurements, a comparison with the electron densities in their figure 2 suggests that 
our spectra in the range of velocities where the metastable transitions are detected traces the outflow at 
distances $\gtrsim 100$ pc from the radiation source. This result is consistent with the recent findings by 
\cite{Arav2018} and those presented by the same group in the works of, e.g., 
\cite{Borguet2013,Xu2018b,Xu2018}.

%-------------------------------------- DISCUSSION ---------------------------------------
\section{Discussion}\label{sec:discussion}

       The potential impact of collisional processes on the radiative acceleration is discussed in \S~\ref{sec:colrad} 
below, and correlations between our results and those from polarization studies are presented 
in \S~\ref{sec:polarization}. We discuss the lack of correlation between the equivalent widths and the 
magnitude of the quasars in \S~\ref{sec:eqwmag}, and highlight some caveats 
and limitations in \S~\ref{sec:caveats}. We conclude by mentioning the applicability of our findings 
for future work in \S~\ref{sec:future}

%-------------------------------------- collisions ---------------------------------
\subsection{Collisions Suppressing Radiative Acceleration?}\label{sec:colrad}

      The important effect of collisions on the species in the outermost UV layer of our 
outflow model might be related to the lack of radiative acceleration from O{\sc vi} and N{\sc v} that we 
inferred in \cite{Masribas2019}. Our results there showed that the absorption 
features of these two ions had a strength similar to that of C{\sc iv}, indicating that their optical depth and in 
turn the amount of absorbed (i.e., scattered) radiation are large, but their line-locking signatures 
revealing acceleration were not detected. It is possible that the radiative acceleration experienced by these 
ions  from the momentum injected via the absorption and emission of photons is  
counterbalanced by Coulomb collisions between the metal ions and the free particles in the gas, 
specifically protons and electrons from ionized hydrogen. 

       \citealt{Baskin2012} quantified this effect in the BLR of AGNs and their fiducial calculations 
indicated that the outflow needs to be exposed to a large photon flux for 
the radiative acceleration to dominate over the collisions. Since the distance from the source is 
larger for the BALs than for the BLR, it is plausible that the collisions suppress the line locking in our case. 
However, if this is the reason for the non-detection of O{\sc vi} and N{\sc v} line locking, the question 
that promptly arises is what prevents C{\sc iv} from suffering the same fate. The temperature of the region 
where C{\sc iv} resides can be certainly lower than that of the two ions, and the density larger than in 
the highly-ionized gas, but these seem to be small differences compared to the largely distinct consequences. 
Future work needs to investigate this processes in more detail, as well as confirming our inferred lack of 
radiative acceleration from the high ions.

%-------------------------------------- polarization ---------------------------------
\subsection{Correlations with Polarization}\label{sec:polarization}

    Polarization observations in BAL quasars and outflows show that there are large differences from 
object to object, but some trends have been observed for the average populations 
\citep[e.g.,][]{Dipompeo2010,Dipompeo2012}. For example, the polarization at the absorption 
troughs is in general larger than that of the quasar continuum and emission lines \citep{Ogle1999}. 
This confirms that scattering of photons in outflow scenarios is important, as scattering polarizes the 
radiation. Furthermore, \cite{Lamy2004} performed correlation and principal component analyses to 
optical spectra and optical polarization data from a set of 139 BAL quasars. They found that BAL troughs 
that start close to the emission lines (small outflow velocities) are more polarized than those that 
start far from the emission (detached), with large detachments implying large angles between the 
line-of-sight toward the observer and the accretion disk. The anti-correlation between polarization and 
detachment was also suggested by \cite{Dipompeo2012}, but these authors argued against the 
viewing angle dependence as the unique reason for the decrease in the polarization. 

     Considering these polarization results and our findings, it is worth noting a potential physical relation   
between the outflow gas density and the degree of polarization. It is possible that the decrease in 
polarization arises from the increase in the optical depth of the outflowing gas with velocity. An optically 
thin medium results in a small number of scattering events, which produces a significant  level of overall 
polarization in the scattered flux. However, when the density increases the photons undergo many more 
scattering events, enabling a broad range of individual polarization values that average to a small total 
degree of polarization \citep[e.g.,][]{Kim2007}. This relation between the column density and the 
polarization in AGNs has been shown in the numerical radiative transfer calculations of \cite{Lee1997} and 
\citealt{Wang2007} (see also \citealt{Wang2005} and \citealt{Chang2015,Chang2017}). 
However, the scenario here suggested implies that the polarization arises from resonant-line scattering 
in the outflow. This is in conflict with some models where polarization is driven by scattering off free 
electrons that inhabit regions separated from the outflowing wind \citep[e.g.,][and references 
thereafter]{Lamy2004}.

%-------------------------------------- EQW vs MAG ---------------------------------------
\subsection{On the Non-correlation of Equivalent Width and Quasar Brightness}\label{sec:eqwmag}

    In \S~\ref{sec:magnitude} we found no correlation between the equivalent widths of the absorption 
features and the absolute quasar magnitude 
(Figure \ref{fig:mag}). This may be not surprising since observations indicate that the brightness of 
the quasars regulates the maximum velocity reachable by the outflows, but not the minimum 
\citep{Ganguly2007}. If the range of magnitudes in our analysis enables similar maximum 
velocities, then our composites could result from samples with the same distribution of 
outflow velocities, and differences due to magnitude would not be visible in their average values. 
Indeed, we find no correlation between outflow velocity and quasar magnitude in our BAL catalog data. 

    The fact that the brightness of the quasars sets only the maximum outflow velocity may support 
the idea that the exact value of the gas velocity depends more on the physical properties of the 
outflow than those of the radiation source, as long as the latter provides enough photon flux to 
radiatively accelerate the material \citep{Ganguly2008}. 

   We conclude by noting also a lack of correlation between equivalent width and X-ray flux that 
has recently been inferred by \cite{Chartas2018} in a narrow-line Seyfert galaxy previously 
analyzed by \cite{Parker2017}.

%-------------------------------------- caveats ---------------------------------------
\subsection{Caveats and Limitations}\label{sec:caveats}

   We have applied the same methodology throughout  to consistently compare our  
stacked spectra. However, potential individual departures from the average 
behavior identified during our analysis are highlighted here. We stress that these 
effects do not impact our current results or interpretations. 

     In the evolution of the equivalent width with trough width (\S~\ref{sec:width})  
some lines appeared shifted from their expected positions. This happens 
in the two composite spectra with the lowest average trough widths. There is no clear dependence 
on the ionization stage, but the fits are slightly better if we consider for these lines an offset of $30-
40\,{\rm km\,s^{-1}}$ from the others (note that our pixel resolution is of $\sim 70\,{\rm km\,
s^{-1}}$).  In the same spectra, the absorption features of N{\sc v} and O{\sc vi} appear narrower 
than that of Ly$\alpha$, and Si{\sc iv} shows width values in between. These differences 
disappear, and all transitions have a similar width, at larger trough width bins.  
In the evolution with outflow velocity (\S~\ref{sec:velocity}) and for the largest velocity bins, 
the width of N{\sc v} and O{\sc vi} seems to be larger than for other ions. All these effects 
are likely due to the different impact of collisions on the different ions, which needs to be 
assessed in more detail in future work.  

   We emphasize the similar shape between the evolution of the C{\sc iv} equivalent width 
with outflow velocity ({\it bottom right panel} of Figure \ref{fig:velocity}) and the distribution of 
the number of C{\sc iv} troughs with outflow velocity in the BAL catalog 
({\it upper panel} of Figure 2 in \citealt{Masribas2019}). We believe that 
the velocity distribution of absorption troughs in the catalog may be driven (biased) by the observability of the troughs, given the 
dependence of the strength of the absorption features on velocity. This distribution would 
simply reflect the difficulty of detecting C{\sc iv} at a given velocity, but it would not represent the 
{\it actual} number of absorbers at that velocity value. In other words, we cannot know 
how many absorbers inhabit each velocity bin since their detectability also depends on velocity.

%-------------------------------------- FUTURE WORK ---------------------------------------
\subsection{Future Work}\label{sec:future}

   We briefly discuss here the impact and possible applications of our results on future outflow studies,  
as well as analyses that can be performed by using our  public composite spectra. 

\begin{enumerate}[leftmargin=0pt,itemindent=20pt]

\item {\it Photo- and collisional-ionization modeling}. These calculations are necessary to measure 
column densities and metallicites, and subsequently obtain more refined values for the densities, 
distances between the different parts of the outflows and the sources, and the energetics and mass 
loading in the outflows. 

\item {\it Detailed absorption-line fitting}. Following the discussion in the previous 
section, a detailed examination of the differences between the width and the offset of the absorption 
features from different ions could reveal important kinematic information for various phases and species, 
and correlations between the kinematic and physical outflow/quasar properties. These differences would 
also yield more information about the impact of collisions on different ions. \\ 
A careful absorption-line search may also result in the detection of molecular hydrogen ${\rm H}_2$. 
Since we estimated outflow distances of a few hundreds of pc from the radiation sources, this detection  
would be of great importance for constraining the connection between the quasar outflows and AGN 
feedback \citep[e.g.,][]{King2015,Harrison2018}. 

\item {\it Observations}. We showed that the largest equivalent widths 
for high-ionization transitions occur in outflows with velocities of a few thousands of  ${\rm km\,s^{-1}}$, 
and below this value for the low-ionization species on average. Below $\sim 500\,{\rm km\,s^{-1}}$ and 
above $\sim 10^4\,{\rm km\,s^{-1}}$, the observations might be contaminated by the host galaxy.  
The equivalent widths of both high- and low-ionization 
lines increase with trough width, therefore broader troughs will typically show stronger features  
and will enable  detecting the faintest associated absorption components. This may be important, for 
example, for the search and accurate measurements of dust indicatiors, e.g., 
absorption features from chromium, and for informing observations that search for molecular gas 
in outflows, such as those currently performed with the Atacama Large Millimeter Array 
(ALMA; e.g., \citealt{Garcia2016,Bonzini2017,Imanishi2018,Izumi2018,Spilker2018}).

\item {\it Radiation and magnetohydrodynamic simulations}. An important byproduct from 
our work is that our results can be used as a bench test for numerical simulations. 
Modern simulations account for many of the physical processes involved in quasar outflows, but 
the lack of constraining observational information enables a large number of plausible simulated scenarios 
\citep[e.g.,][]{Ciotti2010,Choi2015,Weinberger2017}.  Furthermore, most 
simulations focus either on the processes at scales of the order of a parsec around the black holes 
\citep[e.g.,][]{Williamson2018} or at scales beyond the kpc where the outflows interact with the 
interstellar medium. The same computations also typically concentrate their numerical power either 
on radiative transfer or hydrodynamical processes, and just recently they started incorporating 
simultaneously both aspects \citep[][and references therein]{Higginbottom2013,Matthews2016,Dyda2017,Waters2017,Dannen2018} and 
considering a broad range of spatial scales \citep[see the detailed introduction by][]{Barnes2018}. 
The results presented in our current work, as well as in \cite{Masribas2019}, 
concern the outflowing media inhabiting the spatial scales in between those just mentioned, 
and where the coupling between radiation and matter is of great importance. Therefore, the 
advance of numerical simulations in these regimes informed by our results can constrain the 
number of valid simulated alternatives and bridge the gap between quasar outflows and 
AGN feedback. Reciprocally, simulations are also needed to accurately test our interpretations 
and predictions, such as as those for polarization, radiation shields, orientation, shocks and 
radiation pressure, etc., as well as those from other studies.

\end{enumerate}

%-------------------------------------- CONCLUSIONS ---------------------------------------
\section{Conclusion}\label{sec:conclusions}

   We have analyzed  66 quasar outflow composite spectra that extend the 36 previously 
computed for the study of line-locking signatures in \cite{Masribas2019}. The composites 
are built from splitting the overall absorption trough data from the twelfth data release of the SDSS-III/BOSS 
quasar catalog in subsets of outflow velocity, 
width of the absorption troughs, detachment between the absorption troughs and the emission sources, 
and quasar magnitude and redshift. We have measured the equivalent widths of the line and line-locking 
components of more than 100 absorption features in each spectrum (\S~\ref{sec:methods}), and have 
assessed the dependences 
of the equivalent widths on the aforementioned outflow and quasar parameters (\S~\ref{sec:dependencies}). 

The 36 outflow composites and the atomic data for absorption lines are publicly available 
at \url{https://github.com/lluism/BALs}.  

Our findings can be summarized as follows:

\begin{itemize}

\item The outflow absorption spectra are characterized by strong features of high-ionization species    
	but those of low-ionization elements are also present in all cases. Despite the clear detection of 
	many low (neutral) ions, we do not  
	identify features of molecular hydrogen, ${\rm H_2}$, although these might be masked by the 
	line-locking components  of the absorption lines. All high-ionization absorption features are stronger 
	in the overall outflow  
         composite than in that of DLAs, except for the features of the Si{\sc iv} doublet 
         (\S~\ref{sec:dissection} and Figure 
         \ref{fig:stack}).
          
      \item In general, the equivalent widths of both low- and high-ionization absorption features remain fairly 
      constant with outflow velocity in the first $\sim 2\,500\,{\rm km\,s^{-1}}$, and decrease rapidly at higher velocities.
      However, the transitions of the O{\sc vi} and N{\sc v} doublets show a different behavior: their 
      equivalent 
      widths raise steeply with outflow velocity and peak at  $\sim 2\,500\,{\rm km\,s^{-1}}$ before the general 
      decrease. The evolution of these two ions matches that of the parameter we use to estimate the width 
      of the absorption lines (Figure \ref{fig:velocity}).
      
      \item The equivalent width of all the species increases with trough width, but the evolution is the steepest for 
      O{\sc vi} and N{\sc v} (Figure \ref{fig:width}).
      
      \item The maximum difference between the equivalent widths of O{\sc vi} and N{\sc v} and  
      those of other ions occurs at outflow velocities of a couple of thousands of ${\rm km\,s^{-1}}$ and for 
      absorption troughs of similar width values. The outflow velocity bins $\le 800\,{\rm km\,s^{-1}}$ and 
      $\ge 14\,000\,{\rm km\,s^{-1}}$ show  little equivalent 
      width evolution with trough width, suggesting that these velocities may trace media other than the outflow, 
      i.e., the host halo gas, and/or a transition between two physically distinct gas regions (Figure \ref{fig:big}).
           
      \item We find no correlation between the equivalent widths and quasar absolute magnitude or quasar redshift, 
      consistent with the findings from other works (\S~\ref{sec:magnitude}).  
      
       \item The equivalent width ratios between the two lines of high-ionization atomic doublets show 
      average values around $1.25$ in all the subsamples, implying that the composites consist of a combination 
      of strongly and mildly saturated  lines. 
      
      \item We propose a multiphase stratified `Russian-doll' structure for the outflows, where deeper layers are 
      denser, colder and less affected by the radiation field. This structure enables us to 
      consistently explain all the aforementioned differences observed for the equivalent width evolution of 
      the different ions, as well as the behavior of Si{\sc iv} in quasar outflows and DLAs. It also 
      accommodates naturally 
      the origin of the radiation shield that is often invoked for radiative acceleration in outflows  (Figure \ref{fig:flow} 
      and \S\S~\ref{sec:doll}, \ref{sec:shield}). 
      
      \item Detections of transitions from collisionally-populated metastable excited states in some of the outflow 
      velocity stacks reveal the 
      presence of collisional processes. We argue that collisions are the drivers of  the equivalent width 
      differences observed between ions as the strength of these metastable-level features correlates with 
      the equivalent widths of O{\sc vi} and N{\sc v} and the average width of the absorption lines 
      (\S~\ref{sec:collis} and Figure \ref{fig:sivcoll}). 
      
      \item We also infer a correlation between the electron density, the strength of the metastable-level transitions  
     and the outflow velocity, which we interpret as the signature of pressure waves that sweep up and compress 
     material outwards along with the movement of the outflow. Given the ubiquity of the radiative-acceleration 
     signature found in \cite{Masribas2019} and the similar evolutions observed here for the line and 
     line-locking absorption components, we argue that radiation pressure is the main driver of the waves.      
     The pressure-wave scenario fits well in our proposed outflow structure and it can explain the formation and 
     survival of low-ionization species in dense parts of the wave fronts (\S\S~\ref{sec:collis}, \ref{sec:waves}). 
       
       \item We estimate that outflow gas with velocities of $\sim 2\,000-4\,000\,{\rm km\,s^{-1}}$ traces  
       distances of above $100$ pc from the radiation sources.
            
\end{itemize}

   Our proposed outflow structure can qualitatively account for the differences that we have observed in the 
equivalent width evolutions, but it may have implications in other results. For instance, we discussed the 
possible relation between the increase of density and the decrease of polarization in the absorption 
troughs  with velocity (\S~\ref{sec:polarization}). Furthermore, the collisional scenario could be related to 
the lack of line-locking signatures from doublets other than C{\sc iv} that we inferred in 
\citealt{Masribas2019} (\S~\ref{sec:colrad}). 
However, these and other interpretations need to be investigated in more detail in future work. For this 
purpose we have made our main composite spectra publicly available, and we have highlighted  
aspects that can be addressed with photo- and collisional-ionization modeling, and radiation and 
hydrodynamic simulations (\S~\ref{sec:future}). Finally, our results can also 
inform observational campaigns that target outflows, indicating what properties of the outflows are 
connected to the presence or strength of certain species.

%-------------------------------------- ACKNOWLEDGEMENTS ---------------------------------------
\section*{acknowledgements}

I thank the referee for an in-depth and constructive revision of the manuscript that has 
improved the presentation and analysis of the results. 
The initial inspiration for this work grew out of a stimulating discussion with Paul 
Martini during a visit supported by the Visitor Program at the Ohio State Center
for Cosmology and Astroparticle Physics. I am grateful to him for valuable 
ideas and comments on this manuscript, and to the CCAP for kind hospitality. Thanks 
also to Renate Mauland for her work on the spectral stacks in this and our previous paper 
and for several revisions of this manuscript. I am thankful to Ski Antonucci for enriching 
and valuable discussions on polarization in outflows, Tzu-Ching Chang, Robert Wissing, 
Phil Berger, Sijing Shen and Joop Schaye for conversations on shocks, waves and the 
outflow structure, and Max Gronke for discussions on gas hydrodynamics. Thanks to Daniel 
Proga for a careful review of this manuscript and for providing many great comments, 
suggestions and references, and Stan Owocki for sharing his thoughts and notes on hydrostatic 
cloud modeling. I am grateful to Suoqing Ji, Fred Hamann, Sterl Phinney, Nick Scoville, Eliot 
Quataeret, Tom Barlow and other colleagues at JPL and Caltech for many interesting comments 
and discussions. I am thankful to the UCSB/MPIA ENIGMA group, for their kind hospitality and 
comments. This research was partially carried out at the Jet Propulsion Laboratory, California 
Institute of Technology, under a contract with the National Aeronautics and Space Administration.

   Funding for SDSS-III has been provided by the Alfred P. Sloan Foundation, 
the Participating Institutions, the National Science Foundation, and the U.S. 
Department of Energy Office of Science. The SDSS-III web site is 
\url{http://www.sdss3.org/}. SDSS-III is managed by the Astrophysical Research
Consortium for the Participating Institutions of the SDSS-III Collaboration 
including the University of Arizona, the Brazilian Participation Group, Brookhaven 
National Laboratory, Carnegie Mellon University, University of Florida, the French 
Participation Group, the German Participation Group, Harvard University, the 
Instituto de Astrofisica de Canarias, the Michigan State/Notre Dame/JINA 
Participation Group, Johns Hopkins University, Lawrence Berkeley National 
Laboratory, Max Planck Institute for Astrophysics, Max Planck Institute for 
Extraterrestrial Physics, New Mexico State University, New York University, 
Ohio State University, Pennsylvania State University, University of Portsmouth, 
Princeton University, the Spanish Participation Group, University of Tokyo, 
University of Utah, Vanderbilt University, University of Virginia,
University of Washington, and Yale University. 

%-----------------------BIBLIOGRAPHY ----------------------------
%\clearpage
%\bibliographystyle{mn2e}
\bibliographystyle{apj}
\bibliography{bal}\label{References}
%\bibliographystyle{apalike}

%-------------------APPENDIX-------------------------------------------------

\appendix
\clearpage

\begin{figure*}\center %------------  MIN VELOCITY
\includegraphics[width=1.\textwidth]{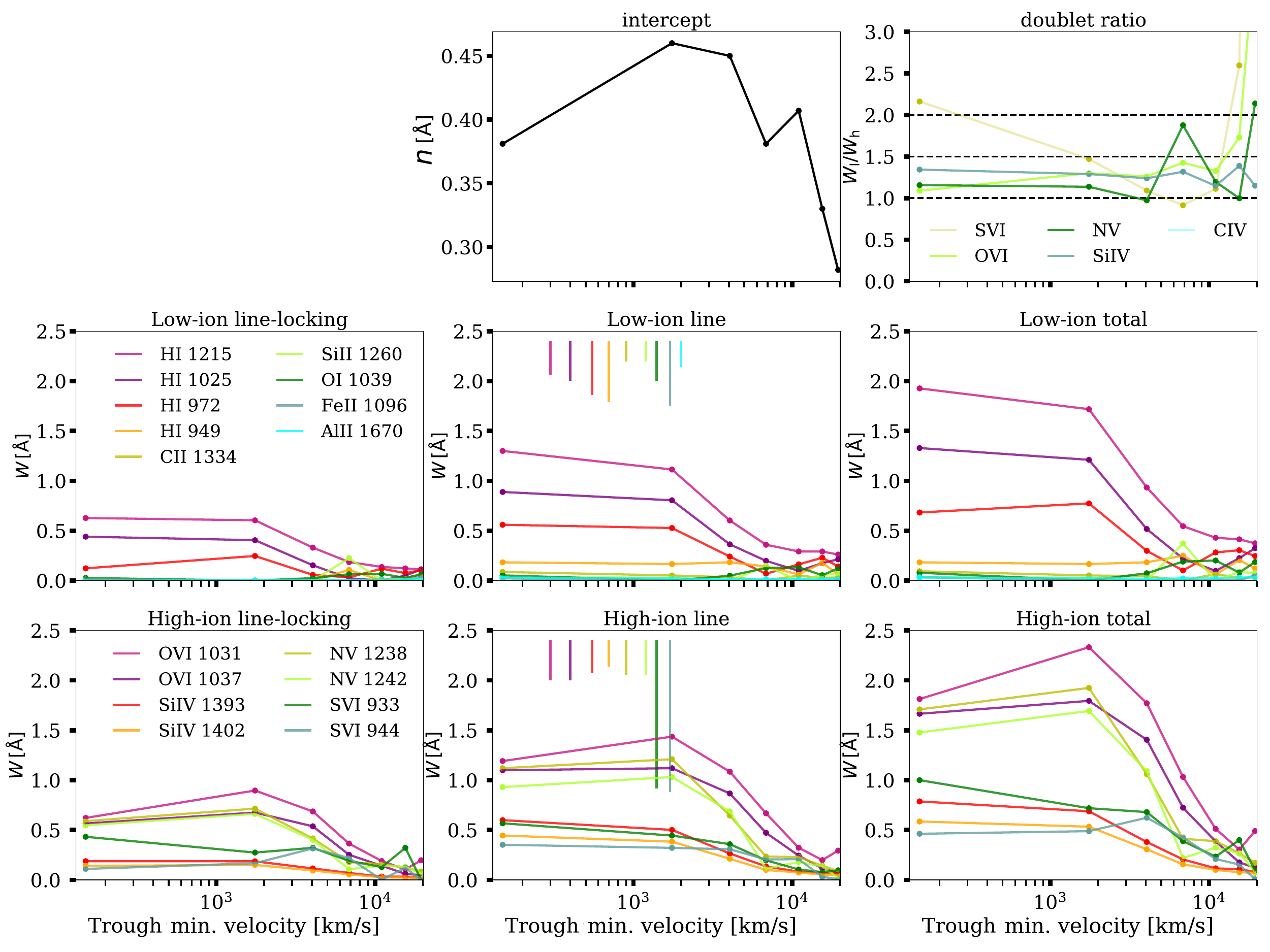}
\caption{Evolution of the equivalent width as a function of trough minimum velocity, following the format of Figure \ref{fig:velocity}. There are no apparent differences beyond the uncertainty between troughs detached or not (smallest velocity bin) from the C{\sc iv} emission line, other than the evolution already observed with outflow velocity.}
\label{fig:vmin}
\end{figure*}

\begin{figure*}\center %------------  MAG
\includegraphics[width=1.\textwidth]{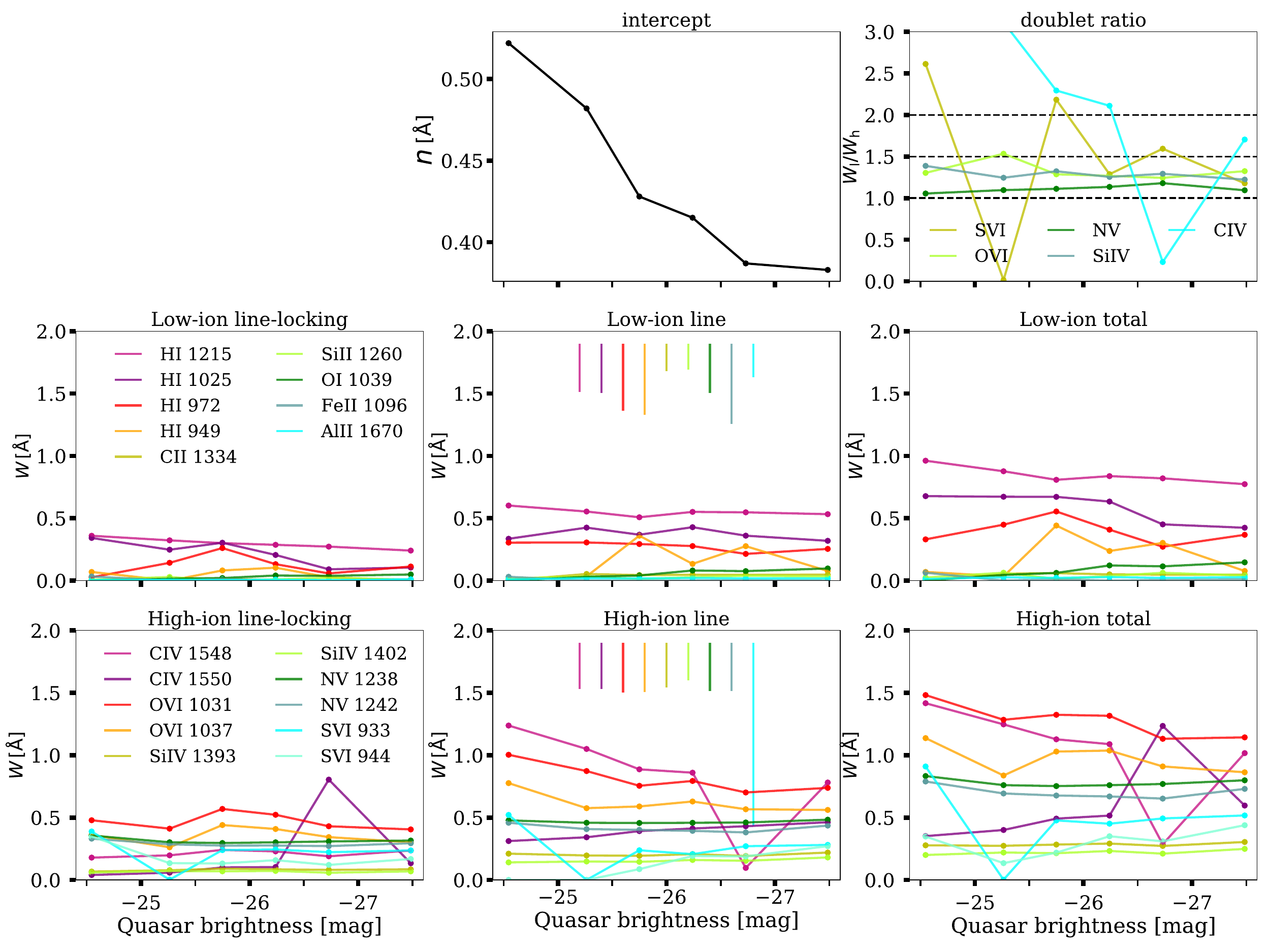}
\caption{Evolution of the equivalent width as a function of quasar absolute magnitude, following the format of Figure \ref{fig:velocity}. There is no visible dependence of the equivalent width on the magnitude of the host quasar beyond the uncertainties. For some high-ionization species, a decrease of equivalent width with increasing brightness, consistent with the trend of the intercept, is suggested, albeit also within 
the uncertainty.}
\label{fig:mag}
\end{figure*}

\begin{figure*}\center %------------  REDSHIFT
\includegraphics[width=1.\textwidth]{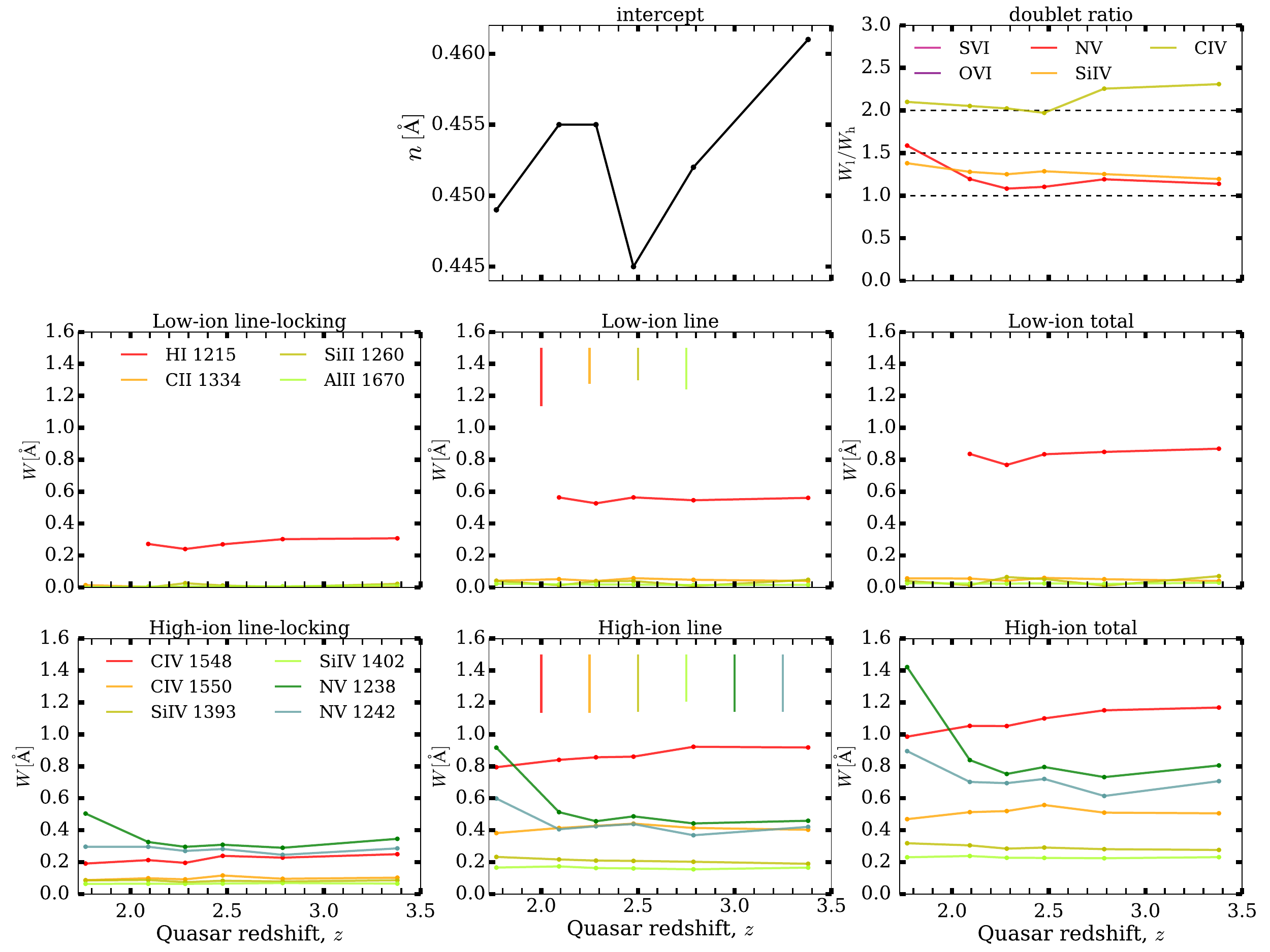}
\caption{Evolution of the equivalent width as a function of redshift, following the format of Figure \ref{fig:velocity}. In general, the measured equivalent widths remain constant with redshift, indicating that the local conditions of the radiation field and the physical state of the gas govern the observations.}
\label{fig:redshift}
\end{figure*}

\clearpage
%-------------------------------------- SPEARMAN RANK CORRELATION ANALYSIS -------------
\section{Spearman Rank Correlation Analysis}\label{sec:spear}

\begin{figure*}\center %------------  REDSHIFT
\includegraphics[width=0.6\textwidth]{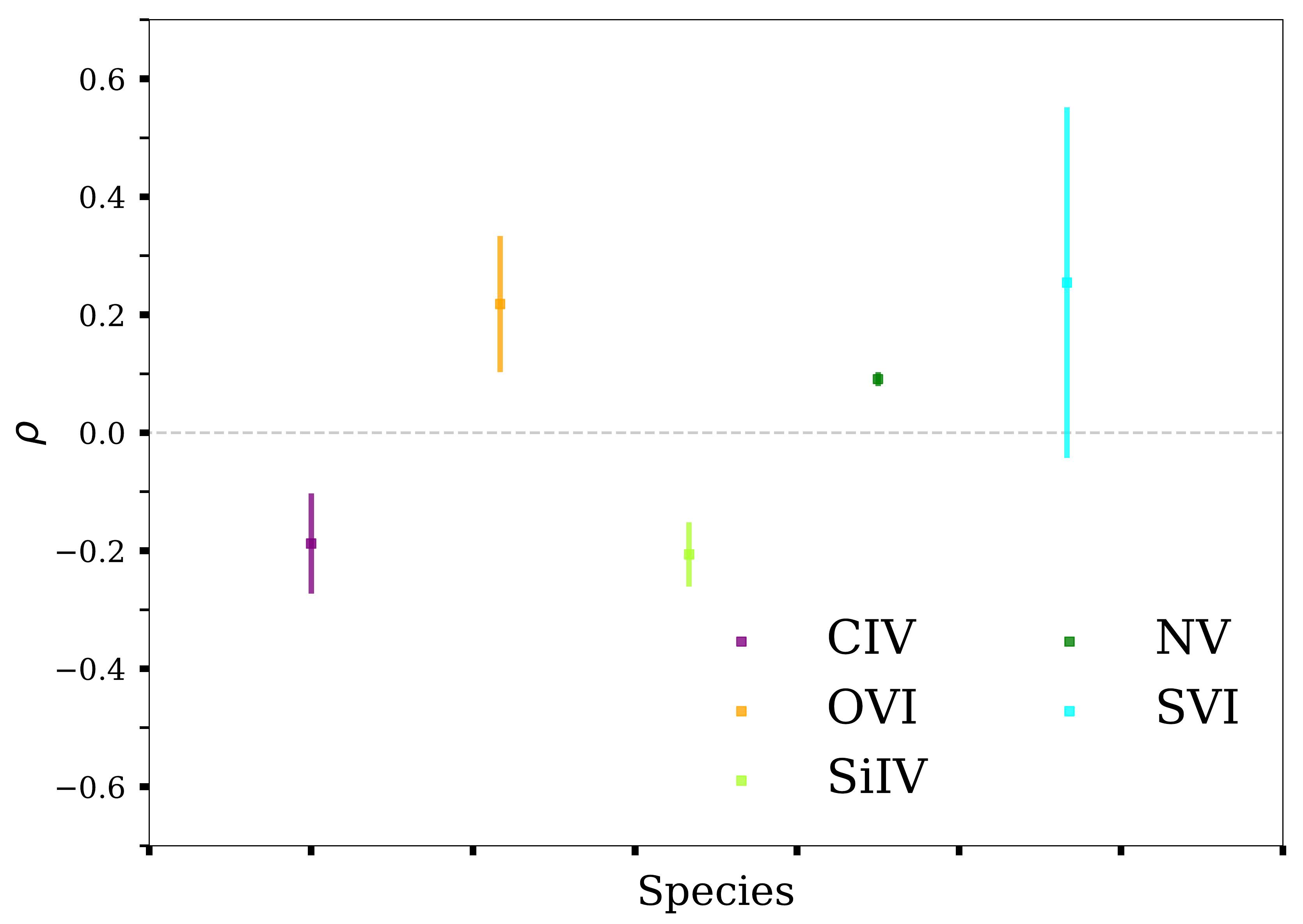}
\caption{Values of the Spearman rank correlation analysis for the high-ionization species and the 
intercept with outflow velocity. Positive (negative) values of the correlation coefficient indicate 
correlation (anti-correlation) between the evolution of the high ions and the intercept, while a null 
value denotes no correlation. The species O{\sc vi}, N{\sc v}  and S{\sc vi} present positive correlations, 
the latter with large error bars. See text for further details.}
\label{fig:rank}
\end{figure*}

    We quantify the correlation between the high-ionization species and the intercept `n' for the 
evolution with outflow velocity (Figure \ref{fig:velocity}). Figure \ref{fig:rank} shows the results of 
a Spearman rank correlation analysis for these high ions. The values and uncertainties are the 
mean value and the standard deviation  from the two lines of each doublet. The species C{\sc iv} and 
Si{\sc iv} present negative values of the correlation coefficient, which indicates an anti-correlation 
between their evolution and that of the intercept. A positive correlation is observed for O{\sc vi} and 
N{\sc v} as suggested in \S~\ref{sec:bump}, although the correlation coefficient value is small, 
$\approx 0.2$ and $\approx 0.1$, respectively. The average value for S{\sc vi} is also positive, 
$\approx 0.25$, but with large error bars consistent with no correlation ($\rho=0$) due to the 
large differences in the evolution of the two lines of the doublet at small velocities ({\it bottom panels} 
in Figure \ref{fig:velocity}). 

\end{document}